\documentclass[12pt,a4paper]{article}

\newif\ifdraft
\draftfalse     

\usepackage[a4paper,margin=2.3cm]{geometry}
\usepackage[british]{babel}
\usepackage[T1]{fontenc}
\usepackage[utf8]{inputenc}
\usepackage{lmodern}
\usepackage{microtype}
\usepackage{csquotes}
\usepackage[normalem]{ulem}

\ifdraft
\fi

\usepackage{mathtools,amssymb,amsfonts,bm}
\usepackage[version=4]{mhchem}
\usepackage{siunitx}

\newcommand{\pdt}[1]{\frac{\partial #1}{\partial t}}

\newcommand{\BeFfour}{\ce{BeF4^{2-}}}
\newcommand{\Tplus}{\ce{T+}}
\newcommand{\TF}{\ce{TF}}
\newcommand{\HT}{\ce{HT}}
\newcommand{\Ttwo}{\ce{T2}}

\usepackage{graphicx}
\usepackage{xcolor}

\usepackage{booktabs,tabularx,array,makecell,longtable}
\newcolumntype{Y}{>{\raggedright\arraybackslash}X}
\newcolumntype{P}[1]{>{\raggedright\arraybackslash}p{#1}}

\usepackage{caption}
\usepackage[section]{placeins}

\ifdraft
  \usepackage[left,mathlines]{lineno}
  \modulolinenumbers[1]

  \newcommand{\draftnote}[2]{%
    \par\smallskip\noindent
    \begingroup
    \setlength{\fboxsep}{5pt}%
    \colorbox{#1}{%
      \parbox{\dimexpr\linewidth-2\fboxsep\relax}{#2}%
    }%
    \endgroup
    \par\smallskip
  }
\else
  \newcommand{\draftnote}[2]{}
\fi

\usepackage{authblk}

\usepackage[numbers,sort&compress]{natbib}
\usepackage{xurl}
\usepackage[hidelinks]{hyperref}
\usepackage{bookmark}
\usepackage[nameinlink,noabbrev]{cleveref}
\crefname{equation}{Eq.}{Eqs.}
\Crefname{equation}{Equation}{Equations}
\crefname{figure}{figure}{figures}
\Crefname{figure}{Figure}{Figures}
\crefname{table}{table}{tables}
\Crefname{table}{Table}{Tables}
\crefname{section}{Section}{Sections}
\Crefname{section}{Section}{Sections}

\newcommand{\papertitle}{Tritium speciation and transport in molten FLiBe: parameters, uncertainties and validation priorities}
\newcommand{\paperpdftitle}{\papertitle}
\newcommand{\paperauthors}{Joe Lunn, Teresa Orr, Megan Leyland, Ayobami Daramola, Rémi Delaporte-Mathurin, James Dark, Graeme J. Ackland, Prashanth Srinivasan, Andrew P. Horsfield, Mikhail Yu. Lavrentiev, Duc Nguyen-Manh, Mark R. Gilbert and Matthew R. Ryder}

\title{\bfseries \papertitle}

\author[1]{Joe Lunn}
\author[1]{Teresa Orr}
\author[1]{Megan Leyland}
\author[1]{Ayobami Daramola}
\author[2]{Rémi Delaporte-Mathurin}
\author[2]{James Dark}
\author[3]{Graeme J. Ackland}
\author[1]{Prashanth Srinivasan}
\author[1]{Andrew P. Horsfield}
\author[1]{Mikhail Yu. Lavrentiev}
\author[1]{Duc Nguyen-Manh}
\author[1]{Mark R. Gilbert}
\author[1]{Matthew R. Ryder\thanks{Corresponding author: \href{mailto:matthew.ryder@ukaea.uk}{matthew.ryder@ukaea.uk}}}

\affil[1]{United Kingdom Atomic Energy Authority, Culham Campus, Abingdon, Oxfordshire OX14 3DB, United Kingdom}
\affil[2]{Plasma Science and Fusion Center, Massachusetts Institute of Technology, Cambridge, MA 02139, USA}
\affil[3]{Centre for Science at Extreme Conditions, School of Physics and Astronomy, University of Edinburgh, Edinburgh EH9 3FD, United Kingdom}

\date{}

\hypersetup{
  pdftitle={\paperpdftitle},
  pdfauthor={\paperauthors}
}

\newcommand{\keywords}[1]{\par\bigskip\noindent\textbf{Keywords:} #1\par}

\begin{document}

\maketitle
\ifdraft\linenumbers\fi

\begin{abstract}
Molten lithium fluoride--beryllium fluoride (FLiBe) is a candidate liquid breeder material for fusion applications, but predictive modelling of tritium retention, transport and extraction depends on chemical and interfacial assumptions that remain only partly resolved. Tritium produced in lithium-bearing fluoride salts may occur in oxidised or fluoride-associated states and can exchange into molecular hydrogen-isotope species such as \HT{}. The relative populations and apparent transport behaviour depend on salt composition and condition, redox state, and interactions at gas--liquid and salt--wall boundaries. This review assesses the evidence for tritium speciation and transport in molten FLiBe, with emphasis on the solubility, effective diffusivity, permeability, exchange and boundary-transfer parameters required by continuum models. The available evidence motivates a two-pool continuum description with separate oxidised or fluoride-associated and molecular-hydrogen pools, without assigning a unique microscopic species. This description applies where states omitted from these pools do not materially affect tritium inventory or transport in the salt. We distinguish directly measured quantities from fitted or inferred effective parameters, recommend how source-specific transport data should be reported and compared, and identify the measurements required to constrain the remaining uncertainties in speciation, exchange, bulk transport and boundary transfer.
\end{abstract}


\section{Introduction}\label{sec:introduction}

Fusion breeding blankets must generate, retain, extract and account for tritium under coupled irradiation, thermal, chemical and material-boundary conditions. Molten lithium fluoride--beryllium fluoride (FLiBe) is an attractive candidate liquid breeder because lithium provides the tritium-breeding reactions, beryllium contributes neutron multiplication, and the salt has favourable high-temperature properties. Tritium behaviour nevertheless remains a coupled problem involving chemical speciation, bulk transport, interfacial mass transfer and structural permeation rather than a simple source-and-diffusion problem \citep{forsberg2017tritiumControl,forsberg2020flibeBlankets,meier2014blankets,pint2024materials,barker2025flibe}.

For predictive modelling and the design of tritium extraction systems, the central issue is not only how much tritium is produced, but in what chemical form or effective transport pool it exists, how rapidly each pool moves through the salt, how the pools are coupled to each other, and how tritium partitions at gas, structural and extraction boundaries. These questions are complicated by salt composition, purification history, oxide and hydroxide impurities, fluorine potential, neutron-induced transmutation products, container interactions and redox-control strategies \citep{shaffer1971preparation,simpson2006redox,zhang2018redox,vergari2023corrosionActivation}.

The aim of this review is to establish which assumptions about tritium speciation and transport in molten FLiBe are supported by the available evidence, where important uncertainties remain, and which measurements would most improve predictive breeder-blanket models. The discussion focuses on oxidised or fluoride-associated tritium, commonly represented in the literature as \Tplus{} or \TF{}, and neutral molecular tritium, represented principally as \HT{} under hydrogen-containing extraction conditions. Their implications for inventory, release, permeation and experimental validation are assessed explicitly.

The scope is intentionally narrower than a general review of FLiBe chemistry or blanket design. Thermophysical properties, neutronics, corrosion and tritium accountancy are considered only where they determine the chemical state, transport parameter, source term or boundary condition used in a predictive model. The objective is not to assign a unique microscopic species to every condition, but to identify the minimum physically defensible model, the parameter ranges it requires and the evidence needed to validate it. We distinguish direct measurements from fitted or inferred parameters, computational predictions and modelling assumptions. Hydrogen and deuterium measurements, together with surrogate-salt experiments, provide useful supporting evidence but cannot replace direct measurements of tritium transport in FLiBe.

\section{FLiBe chemistry and tritium speciation}\label{sec:chemistry-speciation}

\subsection{Salt composition, impurities and redox condition}\label{subsec:salt-condition}

\begin{figure}[htbp]
  \centering
  \includegraphics[width=\textwidth,trim=3pt 0 1pt 0,clip]{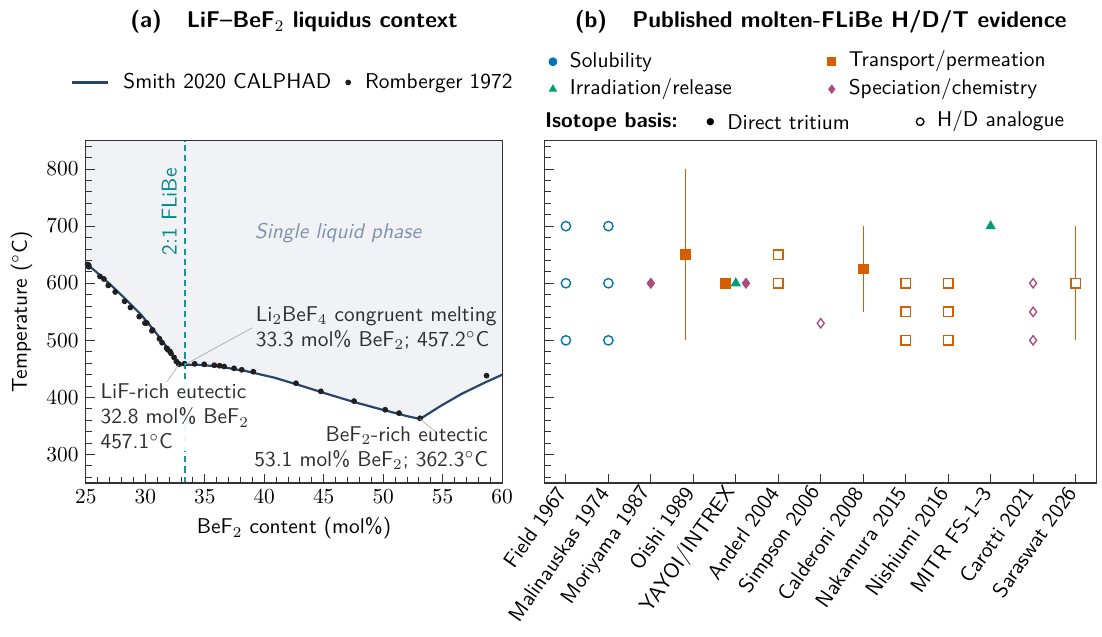}
  \caption{Composition--temperature context and selected published experimental evidence for hydrogen-isotope behaviour in molten FLiBe. \textbf{(a)} LiF--BeF$_2$ liquidus context around nominal 2:1 FLiBe. The Smith et al. CALPHAD liquidus is reconstructed from the published curve and anchored to the calculated invariant temperatures labelled in the panel \citep{smith2020phase,smith2020erratum}. Pale blue shading denotes the single-liquid field above the assessed liquidus; detailed sub-liquidus phase fields are omitted for clarity. Black markers show the Romberger experimental liquidus series \citep{ornl4548,romberger1972liquidus}. The dashed vertical line marks nominal 2:1 LiF:BeF$_2$ (33.3~mol\% BeF$_2$). \textbf{(b)} Selected published experiments directly involving molten LiF--BeF$_2$, grouped by study or campaign and classified by primary evidence type and isotope basis \citep{field1967solubilities,malinauskas1974solubilities,moriyama1987chemical,oishi1989tritiumRecovery,simpson2006redox,calderoni2008permeation,nakamura2015hydrogen,nishiumi2016hydrogen,carotti2021hydrogen,saraswat2026permeation}. Vertical lines denote the temperature spans represented by the plotted measurements, while individual symbols denote discrete temperatures; for range records, the centred symbol identifies the evidence class and isotope basis rather than an additional measurement temperature. Filled symbols denote direct-tritium experiments and open symbols H/D analogues. In the YAYOI/INTREX entry, three horizontally offset symbols distinguish release, speciation and permeation studies from RTNS-II and YAYOI \citep{terai1990ctr,terai1996intrex,terai2001release,suzuki1998speciesChange,suzuki2000permeation,nishimura2001chemical}. The Anderl 2004 markers show the JUPITER-II deuterium permeation measurements \citep{anderl2004deuterium}, with related campaign and analysis papers \citep{fukada2002initial,fukada2003flibeD2,fukada2005diffusion}. The MIT Reactor (MITR) entry groups the FS-1--3 in-core irradiation studies \citep{carpenter2017partitioning,dolan2020graphite,dolan2021tritium}. The remaining entries are labelled by first author and publication year.}
  \label{fig:flibe-composition}
\end{figure}

The FLiBe composition most commonly considered for fusion breeding blankets is a 2:1 molar mixture of \ce{LiF} and \ce{BeF2}, corresponding to \ce{Li2BeF4} (66.7 mol\% \ce{LiF} and 33.3 mol\% \ce{BeF2}). This composition melts congruently and lies close to the LiF-rich eutectic at approximately 32.8 mol\% \ce{BeF2}. A second, lower-temperature eutectic occurs near 53.1 mol\% \ce{BeF2} \citep{romberger1972liquidus,smith2020phase,smith2020erratum}. Because the term \enquote{eutectic FLiBe} is used inconsistently in the literature, compositions are stated explicitly throughout this review. The liquidus context and coverage of the experimental evidence in FLiBe are summarised in \cref{fig:flibe-composition}.

The ideal stoichiometry is useful as a reference state, but practical FLiBe contains impurities inherited from precursor salts, handling, purification and container interactions. Water, oxide, hydroxide and sulphate impurities are important because they alter fluorine potential, corrosion behaviour and hydrogen-isotope chemistry \citep{shaffer1971preparation,guo2018corrosion}. Purification commonly involves treatment with hydrogen-containing or fluoride-containing gases, while beryllium addition and contact with structural alloys can further shift the redox condition. Fu et al.\ reported in situ electrochemical monitoring and potentiostatic removal of oxide ions in 66.7:33.3 mol\% LiF--BeF$_2$ \citep{fu2026oxide}. Interactions with container materials can also modify the impurity inventory and interfacial chemistry during high-temperature experiments \citep{pint2026compatibility}. Consequently, a reported tritium solubility, release rate or apparent diffusivity is not fully defined unless the salt preparation, purification history, impurity burden, redox indicator and container material are also reported. Fluoroacidity, describing Lewis acid--base behaviour in fluoride melts, is distinct from redox condition and can influence both solvent and solute behaviour; the two concepts should therefore not be used interchangeably \citep{hunsberger2026fluoroacidity}.

Molten FLiBe has substantial local structure rather than a random distribution of independent ions. Experimental and simulation studies describe beryllium--fluoride coordination through \BeFfour{}-containing motifs and connected coordination environments, with distributions that depend on composition, temperature and salt condition \citep{baral2021temperature,liu2017molecular,fayfar2024complex}. These environments can affect the availability and dynamics of fluoride coordination sites, the stability of hydrogen-bearing complexes and the migration pathways predicted by atomistic simulations.

\subsection{Tritium production under irradiation}\label{subsec:tritium-production}

The primary tritium source in fusion-relevant FLiBe is neutron interaction with lithium isotopes:

\begin{align}
  {}^{6}\mathrm{Li}(n,\alpha)\mathrm{T},\label{eq:li6-tritium}\\
  {}^{7}\mathrm{Li}(n,n\alpha)\mathrm{T}.\label{eq:li7-tritium}
\end{align}

The \ce{^{7}Li} reaction leaves an outgoing neutron available for further tritium breeding. For both reactions, the chemical state of the tritium after thermalisation depends on interactions with the surrounding salt.

The tritium production rate and its spatial distribution can be obtained from neutronics calculations for the actual LiF:BeF$_2$ composition, lithium isotopic composition, neutron spectrum and blanket geometry. Changes in salt composition alter both the lithium number density and the beryllium neutron-multiplication environment. Because composition also affects tritium chemistry, uncertainties in production and subsequent speciation can be correlated \citep{sawan2003transmutation,ferry2023libra,forsberg2020flibeBlankets}.

Irradiation can also alter the salt chemistry over time. Lithium transmutation, helium production, neutron-induced activation or transmutation products, and irradiation-mediated interactions with structural materials can modify the salt composition and corrosion chemistry with exposure \citep{sawan2003transmutation,vergari2023corrosionActivation}. The tritium production rate and salt chemistry may therefore evolve together during long-duration irradiation experiments or blanket operation.

Oxidising conditions associated with a high fluorine potential favour oxidised or fluoride-associated tritium states, commonly represented as \Tplus{} or \TF{}. Hydrogen-containing purge conditions can promote neutral molecular species through isotopic exchange \citep{terai1996intrex,suzuki1998speciesChange,suzuki1998mechanism,nishimura2001chemical,petti2006jupiter}. Strongly reducing salt chemistries may also stabilise reduced or Be--H-associated states, although their importance in molten FLiBe remains poorly constrained \citep{moriyama1987chemical,carotti2021hydrogen,jiang2025solute,jiang2026interfaces}. Local variations in impurity content, corrosion products and salt stoichiometry can perturb the fluorine potential and therefore the resulting tritium speciation.

\subsection{Oxidised, fluoride-associated and molecular tritium}\label{subsec:modelling-species}

Tritium can occupy different chemical states in molten FLiBe, with consequences for its mobility, retention and interactions at material boundaries \citep{lam2021hydrogenValence,wang2022tritiumFpmd,anderl2004deuterium, Daramola2026FLiBe}. For the continuum model introduced in \cref{subsec:minimum-model}, we group these states into two effective transport pools: oxidised or fluoride-associated tritium and molecular hydrogen isotopologues. The oxidised pool includes protonic tritium associated with the fluoride network and species such as \TF{} where appropriate. Here, \enquote{oxidised} refers to the formal oxidation state of tritium, principally $+1$, rather than the presence of oxygen-containing species \citep{carotti2021hydrogen}. The molecular-hydrogen pool represents neutral molecules such as \HT{} and \Ttwo{} that can exchange with the gas phase. These pools allow different transport and extraction behaviour to be represented without assigning a single microscopic species to each.

The notation \Tplus{} is used in experimental and simulation discussions to represent protonic or cationic tritium \citep{suzuki1998speciesChange,lam2021hydrogenValence}. In a molten fluoride, this should normally be interpreted as tritium associated with the fluoride coordination network, not as a persistent bare nucleus. Atomistic studies suggest that hydrogen valence strongly affects local bonding and migration, but the transferability of a calculated self-diffusion coefficient depends on composition, temperature, cell size, electronic-structure method and imposed charge state \citep{klix2006tritium,lam2021hydrogenValence,wang2022tritiumFpmd, Daramola2026FLiBe}.

Reduced tritium states also warrant explicit consideration. Early radiometric work inferred a negative-valence tritium population from analyses of solid \ce{LiF}--\ce{BeF2} samples dissolved in heavy water \citep{moriyama1987chemical,carotti2021hydrogen}. After adding \ce{LiH}, Carotti et al.\ observed a voltammetric response consistent with oxidation of zero-valent hydrogen. They proposed that \ce{BeH2} could account for the high dissolved hydrogen concentration \citep{carotti2021hydrogen}. Atomistic simulations predict free neutral hydrogen atoms and Be--H complexes in reducing FLiBe \citep{jiang2025solute,jiang2026interfaces}. Direct experimental evidence does not yet constrain a separate transport model for analogous reduced tritium states in molten FLiBe. Their omission from the two-pool model is an approximation whose effect on inventory and transport should be assessed under reducing conditions.

\TF{} represents a neutral molecule with tritium in the formal $+1$ oxidation state. It is chemically plausible under oxidising, high-fluorine-potential conditions and is often invoked when interpreting HF or DF analogue data \citep{field1967solubilities,moriyama1987chemical,petti2006jupiter}. However, an experiment that measures total retained tritium, release or permeation does not necessarily distinguish molecular \TF{} from a broader protonic, fluoride-associated population. Where the species assignment is indirect, the model should use an effective oxidised pool unless molecular \TF{} has been independently established.

\HT{} is the best-supported neutral molecular species under hydrogen-containing extraction conditions. Hydrogen addition promotes isotopic exchange, allowing retained oxidised tritium to enter a molecular form that can be removed through the gas phase \citep{oishi1989tritiumRecovery,suzuki1998htRelease,suzuki1998mechanism,terai2001release,edao2009release}. \Ttwo{} may also form \citep{nishimura2001chemical}. Experiments that distinguish only water-soluble and water-insoluble tritium do not determine the separate contributions of \HT{} and \Ttwo{} to the released gas \citep{terai1990ctr}. The interpretation and recommended continuum treatment of these candidate states are summarised in \cref{tab:species}.

\begin{table}[htbp]
\centering
\caption{Candidate tritium-bearing states in molten FLiBe and their recommended treatment in continuum models.}
\label{tab:species}
\footnotesize
\renewcommand{\arraystretch}{1.15}
\begin{tabularx}{\textwidth}{@{}P{0.11\textwidth}P{0.22\textwidth}Y Y@{}}
\toprule
\textbf{Notation} &
\textbf{Physical interpretation} &
\textbf{Likely conditions and evidence} &
\textbf{Recommended modelling treatment} \\
\midrule
\Tplus{} &
Protonic or cationic tritium associated with the fluoride network &
Inferred in post-production analyses \citep{moriyama1987chemical} and favoured under oxidising conditions; explicitly represented in some atomistic simulations \citep{lam2021hydrogenValence} &
Include within an effective oxidised pool unless independent kinetics justify a distinct ionic species \\
\ce{T-} / reduced T &
Negative-valence or otherwise reduced tritium; Be--H/T association may occur under locally reducing conditions &
Radiometric analysis of dissolved solid samples suggests negative-valence tritium \citep{moriyama1987chemical}. Voltammetry after \ce{LiH} addition is consistent with zero-valent hydrogen \citep{carotti2021hydrogen}. Simulations predict Be--H association under reducing conditions \citep{jiang2025solute}. Direct transport data for reduced tritium are sparse &
Assess as a model-form uncertainty; extend the model where reduced-state storage or transport is significant. If its contribution is unconstrained, report that limitation \\
\TF{} &
Neutral hydrogen-fluoride isotopologue &
Plausible under oxidising, high-fluorine-potential conditions; often inferred through HF/DF analogues and exchange behaviour \citep{field1967solubilities,moriyama1987chemical} &
Treat explicitly only where molecular identity is established; otherwise include within the oxidised pool \\
\HT{} &
Neutral molecular hydrogen isotopologue &
Supported under hydrogen exchange and gas-sweep conditions; central to measured release and extraction \citep{suzuki1998htRelease} &
Use \HT{}-specific bulk properties where \HT{} dominates molecular transport in the salt. \\
\Ttwo{} &
Neutral molecular tritium &
\Ttwo{} may contribute to the water-insoluble fraction recovered in release experiments, but these measurements do not distinguish \Ttwo{} from \HT{} \citep{terai1990ctr,nishimura2001chemical}. &
Include where \Ttwo{} contributes appreciably to the dissolved molecular inventory, bulk transport or boundary flux. \\
\bottomrule
\end{tabularx}
\end{table}

\subsection{Isotopic exchange and redox control}\label{subsec:exchange-redox}

A representative net reaction for isotopic exchange with a protium-containing gas is

\begin{equation}
  \ce{TF + H2 <=> HT + HF}
  \label{eq:isotopic-exchange}
\end{equation}

Equation~\ref{eq:isotopic-exchange} uses \TF{} and \HT{} to illustrate exchange between fluoride-associated tritium and molecular hydrogen isotopologues. Corresponding net reactions can be written for other combinations of hydrogen, deuterium and tritium. These overall exchange relations do not establish the elementary reaction mechanisms in the melt. Irradiation and release experiments show that hydrogen-containing purge gases can shift recovered tritium towards \HT{} and accelerate its removal from FLiBe \citep{moriyama1987chemical,oishi1989tritiumRecovery,terai1990ctr,terai1996intrex,suzuki1998htRelease,suzuki1998speciesChange,suzuki1998mechanism,terai2001release,petti2006jupiter,edao2009release}. The 1 L campaign of the Build A Better Yield (BABY) experiment irradiated ClLiF (\ce{LiF}--\ce{LiCl}) with \SI{14}{\mega\electronvolt} D--T neutrons. Hydrogen addition accelerated tritium recovery, and more than 96\% of the recovered tritium was water-insoluble with and without added hydrogen. The authors proposed isotopic exchange with tritium retained on apparatus surfaces to explain the additional outer-vessel release after hydrogen addition \citep{delaporteMathurin2026baby1l}. The effective conversion rate nevertheless depends on salt condition, hydrogen partial pressure, gas--liquid interfacial area, mixing, temperature and the availability of the oxidised tritium population. A single intrinsic first-order rate constant is difficult to justify unless chemical conversion, interfacial transfer and bulk transport have been separated experimentally.

Beryllium addition can shift the redox condition of fluoride salts \citep{hara2006interactions,petti2006jupiter,simpson2006redox,fukada2006control,fukada2007reaction}, while hydrogen addition or sparging can promote isotopic exchange and tritium removal \citep{oishi1989tritiumRecovery,suzuki1998htRelease}. Electrochemical methods may provide a more direct means of probing or controlling hydrogen redox behaviour, although their quantitative tritium kinetics and applicability at fusion scale remain less well established \citep{carotti2021hydrogen}.

\section{Transport properties and mechanisms}\label{sec:transport}

\subsection{Definitions and comparability}\label{subsec:transport-quantities}

Solubility describes equilibrium or effective partitioning into the salt, while diffusivity describes bulk migration. Permeability is used in different ways across the FLiBe literature: liquid permeability may combine dissolution and diffusion, whereas wall permeability describes transport through a solid structural boundary. We retain each source's phase and coefficient convention and treat gas--liquid exchange separately as interfacial mass transfer. These quantities are transferable between experiments only if the chemical species or effective pool, salt condition, coefficient definition and boundary conditions are comparable \citep{malinauskas1974solubilities,field1967solubilities,calderoni2008permeation,anderl2004deuterium,saraswat2026permeation,yang2026multidimensional}.

For molecular hydrogen-isotope dissolution in FLiBe, a Henry-law relation may be written as

\begin{equation}
  c_i = K_{H,i} p_i,
  \label{eq:henry}
\end{equation}

where $c_i$ is the dissolved molecular concentration, $p_i$ is the corresponding gas-phase partial pressure and $K_{H,i}$ is the Henry-law coefficient under a specified convention. The macroscopic partition relation does not establish the microscopic association or dissociation mechanism in the melt. Where a study defines a liquid permeability as the product of diffusivity and a Henry-law solubility or partition coefficient, that combined quantity is retained under the source convention and distinguished explicitly from permeability through the structural wall.

By contrast, atomic hydrogen dissolution in many metallic materials used in permeation measurements is commonly represented by a Sieverts-law relation,

\begin{equation}
  c_X^{\mathrm{metal}} = K_{S,X}\sqrt{p_{X_2}}, \qquad X\in\{\mathrm{H},\mathrm{D},\mathrm{T}\},
  \label{eq:sieverts}
\end{equation}

where $c_X^{\mathrm{metal}}$ is the molar concentration of dissolved atoms of isotope $X$ in the metal, $p_{X_2}$ is the partial pressure of the corresponding molecular gas, and $K_{S,X}$ is the Sieverts coefficient for the specified temperature and metal \citep{candido2026sensor}. On a molar concentration basis, $K_{H,i}$ has units of $\mathrm{mol\,m^{-3}\,Pa^{-1}}$, whereas $K_{S,X}$ has units of $\mathrm{mol\,m^{-3}\,Pa^{-1/2}}$; the two therefore cannot be interconverted by a pressure-independent factor. Different publications also use different concentration bases, standard states and reciprocal coefficient conventions, so values should be retained in their original form before conversion. In integrated metal--molten-salt permeation experiments, however, the measured pressure dependence can reflect interfacial chemistry rather than a unique bulk dissolution law of the salt. In a recent preprint, Delaporte-Mathurin and Dark predict that competing molecular-recombination and fluorination channels can make the apparent pressure exponent depend on interfacial hydrogen loading and salt redox state \citep{delaporteMathurin2026kineticInterface}. For a permeation relation of the form $J_T=(D_{\mathrm{FLiBe},T}K_{\mathrm{FLiBe},T}/L_{\mathrm{FLiBe}})\sqrt{p_{T_2}}$, dimensional consistency requires $K_{\mathrm{FLiBe},T}$ to have units of $\mathrm{mol\,m^{-3}\,Pa^{-1/2}}$ on a molar-flux basis. Calderoni et al.\ report $K_{\mathrm{FLiBe},T}$ in $\mathrm{mol\,m^{-3}\,Pa^{-1}}$, whereas their square-root-pressure flux relation implies $\mathrm{mol\,m^{-3}\,Pa^{-1/2}}$. The numerical fit is retained as reported and distinguished from Henry-law coefficients. At 973~K, with an upstream \ce{T2} partial pressure of about 1.2~kPa and an 8.1~mm salt layer, Eqs.~(2), (6) and (7) of the source give a flux of $2.3\times10^{-8}\ \mathrm{mol\,m^{-2}\,s^{-1}}$. The corresponding flux read from figure~2 of Calderoni et al.\ is about $3.7\times10^{-7}\ \mathrm{mol\,m^{-2}\,s^{-1}}$. The cause of the discrepancy is unclear \citep{calderoni2008permeation}.

In the absence of advection, bulk transport can be represented as

\begin{equation}
  \pdt{c_i} = \nabla \cdot \left(D_i \nabla c_i\right) + R_i,
  \label{eq:fick-reaction}
\end{equation}

where $D_i$ is a diffusivity and $R_i$ captures production, loss or interconversion. Atomistic self-diffusion, tracer diffusion, diffusivities obtained from capillary measurements, fitted release coefficients and diffusivities inferred from permeation are not automatically equivalent. In a reactive salt, a fitted $D_i$ may be an effective parameter that absorbs exchange, trapping or boundary-transfer limitations \citep{fukada2003flibeD2,anderl2004deuterium,fukada2005diffusion,petti2006jupiter,li2022hfDiffusion,wang2022tritiumFpmd}.

Permeability must be defined with equal care. A reported permeation coefficient may combine dissolution into FLiBe, transport through the salt, transfer at the salt--wall interface, diffusion through a structural material, and recombination or desorption on the receiving side. Such data are valuable for integrated model validation but should not be used as a bulk FLiBe property without deconvolution. Conversely, ex-reactor measurements using tritium generated from neutron-irradiated FLiBe to compare permeation through bare and alumina-coated SS316 isolate the structural-barrier contribution and should not be interpreted as measurements of FLiBe transport itself \citep{zheng2019tpb}. Separating the individual transport resistances is model-dependent and can introduce substantial uncertainty, since they are rarely measured independently. Recent hydrogen-isotope measurements demonstrate that salt--metal wetting, membrane coverage by the salt and interfacial gas bubbles can materially affect the apparent permeation response \citep{saraswat2026permeation}. A recent multidimensional FESTIM analysis of the same experimental platform shows that lateral transport, sidewall leakage and external boundary assumptions can affect the permeability inferred from measured fluxes \citep{yang2026multidimensional}.

\subsection{Nature of the chemical bonding}\label{subsec:chemical bonding}

FLiBe has traditionally been modelled as a molten salt of F$^-$, Be$^{2+}$ and Li$^+$ ions. Fu et al.\ identified tetrahedral Be--F coordination by NMR and interpreted their measurements as evidence of a strong covalent contribution to Be--F bonding \citep{fu2023probing}. In simulated liquids, fluoride sharing links Be--F tetrahedra into dimers and larger oligomers \citep{fayfar2024complex}. The BeF$_4^{2-}$ units are well-established structural motifs, but the observation of tetrahedral coordination does not, by itself, distinguish between an ionic packing model and a covalent bonding model. Assigning atomic charges and bond orders requires a choice of electronic-structure analysis. Smith et al.\ reproduced tetrahedral coordination and fluoride-sharing structures with a polarisable ionic potential \citep{smith2020phase}.

DFT-assisted interpretation of the Raman spectra suggests a progression from isolated BeF$_4$ units to larger fluoride-sharing structures as the \ce{BeF2} content increases \citep{li2021raman}. Assessing whether these motifs move as distinct transport species requires evidence about their lifetimes and the correlated motion of the ions.

The charge state of T is also critical. A potential that reproduces bulk FLiBe structure also needs to be tested for tritium--fluoride coordination, molecular stability and mobility. Ab initio molecular dynamics (AIMD) simulations with prescribed protonic and molecular tritium states can estimate self-diffusivities for those states \citep{lam2021hydrogenValence}. Direct simulation of the isotope exchange in \cref{eq:isotopic-exchange} requires a potential that allows the relevant bonds to form and break. Simulations of separate states can inform the pool-specific transport coefficients in \cref{subsec:minimum-model}, with interconversion rates constrained independently.

\subsection{Solubility constraints}\label{subsec:solubility}

Although hydrogen solubility in FLiBe has been studied for more than five decades, there remains considerable uncertainty in the quantitative data. For \ce{H2} in the Henry-law regime, $p_i$ in \cref{eq:henry} corresponds to the gas-phase partial pressure $p_{\mathrm{H_2}}$, and $K_{H,i}$ corresponds to the Henry-law coefficient $K_{H,\mathrm{H_2}}$, relating equilibrium dissolved molar concentration to partial pressure at the specified temperature and salt chemistry. The corresponding relations for \HT{} and \Ttwo{} give the molecular pool's salt-side equilibrium reference concentration on a tritium-atom basis in \cref{eq:gas-equilibrium}. The difference between the actual salt-side concentration and this reference concentration drives the finite-rate transfer in \cref{eq:gas-transfer}.

Early measurements by Field and Shaffer in the 1960s established the uptake of HF and DF in molten fluoride mixtures \citep{field1967solubilities}. Identifying the microscopic species in the melt requires separate evidence. Malinauskas and Richardson reported measurements of molecular hydrogen solubility in molten \ce{Li2BeF4} in 1974 \citep{malinauskas1974solubilities}. Their combined \ce{H2}/\ce{D2} Ostwald correlation increases with temperature over 773--973~K. Comparisons with calculated solubilities require matching the gas and dissolved species, temperature, salt composition and redox conditions, and using consistent concentration, pressure and standard-state conventions. \Cref{tab:model-parameters} gives the values under their source conventions.

\subsection{Neutral molecular transport}\label{subsec:ht-transport}

\HT{} is a well-supported extraction product under protium-containing conditions. Using \HT{}-specific properties for the dissolved molecular pool requires evidence or an explicit assumption that \HT{} dominates molecular transport in the salt. Classical gas-solubility measurements indicate relatively low solubility of molecular hydrogen isotopes in FLiBe, while HF and DF measurements show substantially greater uptake under the conditions studied. Neutron-irradiation, release and permeation experiments also demonstrate that \HT{} can be recovered under hydrogen-containing conditions \citep{malinauskas1974solubilities,field1967solubilities,suzuki1998htRelease,edao2009release,suzuki2000permeation,calderoni2008permeation}. The BABY \SI{100}{mL} experiments, performed with ClLiF as a surrogate breeder salt rather than FLiBe, recovered predominantly water-insoluble tritium (\HT{} or \Ttwo{}) without distinguishing between these isotopologues \citep{delaporteMathurin2025baby}. The subsequent BABY 1L campaign using the same surrogate-salt system showed that hydrogen addition to the carrier gas accelerated release \citep{delaporteMathurin2026baby1l}. These ClLiF experiments benchmark integrated tritium recovery and gas handling. Bulk-salt exchange kinetics and FLiBe transport properties require independent constraints. The BABY experiments also reinforce the need to report sweep-gas composition, salt condition and apparatus materials when interpreting recovered species.

Atomistic calculations also provide evidence on molecular transport. Lam et al.\ calculated faster self-diffusion for neutral molecular tritium than for fluoride-associated tritium, while Wang et al.\ examined both molecular and fluoride-associated forms \citep{lam2021hydrogenValence,wang2022tritiumFpmd}. Using a fixed-charge potential, Daramola et al.\ calculated higher self-diffusivities for \Ttwo{} than for \TF{} or \ce{T-} at the five temperatures studied (873--1273~K) \citep{Daramola2026FLiBe}. Machine-learning molecular dynamics predicts stable, mobile molecular \ce{H2} under the conditions studied \citep{jiang2025solute}. Using these mobility results to estimate \HT{} or \Ttwo{} transport requires confirming the isotope masses used in the simulations and assessing possible isotope effects.

No single parameter set has yet been demonstrated to be transferable across the available experiments. Substantial scatter exists among molecular-hydrogen solubility, diffusivity and permeability estimates obtained from different FLiBe dissolution and permeation experiments \citep{malinauskas1974solubilities,nakamura2015hydrogen,nishiumi2016hydrogen,saraswat2026permeation}. Some experiments introduce molecular hydrogen or deuterium directly, while others generate tritium by irradiation and then infer transport from release or wall permeation. Salt composition, redox condition, surface area, mixing, gas flow and container material can determine whether the observed response is controlled by bulk diffusion, gas--liquid transfer, chemical conversion or wall transport. Accordingly, each dataset should first be reported in terms of the directly measured quantity, with any additional assumptions required to derive $K_{\mathrm{mol}}$, $D_{\mathrm{mol}}$ or a surface-transfer coefficient stated explicitly.

Hydrogen- and deuterium-based measurements can provide useful constraints on \HT{} transport where the chemical state and transport mechanism are demonstrably comparable, but transfer between isotopes should be justified explicitly \citep{malinauskas1974solubilities,nakamura2015hydrogen,nishiumi2016hydrogen,saraswat2026permeation}. Simple mass scaling may provide a useful estimate for diffusion within an otherwise unchanged mechanism, whereas equilibrium partitioning, reactive exchange and interfacial kinetics can exhibit distinct isotope effects for which no universal H/D/T conversion is justified. Analogue-derived values should retain the original isotope identity and remain provisional until tritium-specific measurements under characterised FLiBe conditions are available.

\subsection{Oxidised and fluoride-associated transport}\label{subsec:nonmolecular-transport}

The major unresolved mechanistic question is whether oxidised tritium transport is best represented as molecular \TF{} diffusion, migration of a protonic species between fluoride coordination environments, correlated ion-pair motion, or an effective reaction-diffusion process \citep{petti2006jupiter,klix2006tritium,lam2021hydrogenValence,wang2022tritiumFpmd,li2022hfDiffusion}. These alternative mechanisms imply different bulk mobilities and partitioning relations at gas and structural boundaries. Oishi et al.\ also measured tritium diffusivity in molten FLiBe by a capillary method. That measurement provides a direct-tritium comparison with the diffusivities inferred from permeation \citep{oishi1989tritiumRecovery}.

HF and DF solubility data provide direct information about hydrogen-fluoride isotopologues, but their use for irradiation-produced tritium requires assumptions about species identity and redox condition \citep{field1967solubilities}. Electrochemical measurements probe hydrogen redox chemistry under deliberately modified salt conditions \citep{carotti2021hydrogen}. Atomistic simulations provide complementary mechanistic information but are sensitive to the imposed valence state, composition and redox environment. Early AIMD calculations found an introduced tritium species to remain associated with fluorine over the simulated trajectory \citep{klix2006tritium}. Lam et al.\ calculated tritium self-diffusivities in FLiBe at 973 K of $(6.3\pm1.4)\times10^{-9}\ \mathrm{m^2\,s^{-1}}$ for the neutral molecular state and $(3.0\pm0.5)\times10^{-9}\ \mathrm{m^2\,s^{-1}}$ for the fluoride-associated \Tplus{} state \citep{lam2021hydrogenValence}. Subsequent AIMD and electrochemical studies provided complementary evidence for condition- and species-dependent fluoride-associated hydrogen and tritium transport \citep{wang2022tritiumFpmd,li2022hfDiffusion}. Machine-learning molecular dynamics studies indicate that reduced solutes, including metallic Be, can alter protium speciation while molecular hydrogen remains comparatively stable and mobile \citep{jiang2025solute,jiang2026interfaces}. Machine-learning simulations also predict that hydrogen adsorption and trapping at FLiBe--graphite and FLiBe--alloy interfaces depend on surface structure and chemical state \citep{jiang2026interfaces}. Other calculations have examined temperature-dependent tritium self-diffusion and T--F coordination in FLiBe, including the influence of helium on local tritium coordination \citep{galashev2023structure,galashev2025firstPrinciples}. Atomistic transport models should also be distinguished by the bonding changes they permit. A classical model that constrains tritium to a fixed fluoride partner can describe fluoride-associated motion but cannot capture fluoride-to-fluoride transfer or interconversion with molecular hydrogen \citep{Daramola2026FLiBe}. These computational results should be compared as condition-specific mechanistic evidence; combining them into a single generic tritium diffusivity would obscure differences in species definition, thermodynamic state and simulation methodology. Coordination and self-diffusion results from separate simulations with prescribed protonic and molecular states do not, by themselves, determine their equilibrium population ratio. Predicting that ratio requires relative free energies under consistent thermodynamic conditions, with the relevant chemical and electronic reservoirs specified. Equilibrium speciation also does not directly determine the source partition following irradiation-induced relaxation.

Experimental measurements and atomistic calculations can both help distinguish these microscopic descriptions. A single effective oxidised pool is appropriate where the fluoride-associated configurations can be represented by effective local transport and exchange coefficients at the spatial and temporal scales being modelled. Its solubility or capacity, effective diffusivity and exchange rate should be treated as uncertain, state-dependent parameters. More detailed models that resolve \TF{} explicitly or represent protonic hopping can be tested against experimental data as they become available. Transport data for reduced or Be--H-associated states are even less mature; these states require separate treatment where they materially affect inventory or transport and cannot be represented adequately by the existing pools.

Fixed atomic charges can limit a potential's description of charge redistribution, while training over a narrow range of oxidation states can restrict its use for redox-driven species interconversion \citep{Daramola2026FLiBe}. Charge-aware or otherwise electronically responsive potentials should therefore be tested across the bonding and redox conditions relevant to FLiBe \citep{ko2021chargeTransfer}. Validation should address the quantities being predicted: agreement with bulk salt structure alone does not establish accuracy for hydrogen diffusion, relative species free energies, exchange barriers or interfacial trapping. Diffusion estimates should also report the simulated isotope mass, trajectory length, cell size and sampling uncertainty.

\section{Parameters for continuum modelling}\label{sec:model-parameters}

\subsection{Minimum useful model}\label{subsec:minimum-model}

Resolving each proposed tritium species would require species-specific transport and reaction parameters that are not yet adequately constrained. We therefore adopt a two-pool formulation with a volumetric tritium source $S_T$ on a tritium-atom basis, source partitioning between oxidised and molecular-hydrogen pools, interconversion, advection in flowing salt, pool-dependent diffusion, and boundary conditions for gas transfer and structural loss \citep{stempien2016trident,humrickhouse2020tritiumTransport,brooks2022blankets,ferrero2022arc,delaporteMathurin2024festim,dark2026festimV2,park2026crossScale}. This description assumes that the oxidised and molecular pools account for the tritium inventory and fluxes of interest in the salt. Unresolved configurations within each pool are assumed to admit effective local transport and exchange coefficients at the spatial and temporal scales being modelled. A persistent reduced population or another slow storage process may require additional state variables.

\begin{align}
  \pdt{c_{\mathrm{ox}}} + \nabla \cdot \left(\bm{u}c_{\mathrm{ox}}\right)
  &= \nabla \cdot \left(D_{\mathrm{ox}} \nabla c_{\mathrm{ox}}\right)
  + f_{\mathrm{ox}}S_T - k_{\mathrm{ox}\rightarrow\mathrm{mol}} c_{\mathrm{ox}}
  + k_{\mathrm{mol}\rightarrow\mathrm{ox}} c_{\mathrm{mol}},\label{eq:model-oxidised}\\
  \pdt{c_{\mathrm{mol}}} + \nabla \cdot \left(\bm{u}c_{\mathrm{mol}}\right)
  &= \nabla \cdot \left(D_{\mathrm{mol}} \nabla c_{\mathrm{mol}}\right)
  + \left(1-f_{\mathrm{ox}}\right)S_T
  + k_{\mathrm{ox}\rightarrow\mathrm{mol}} c_{\mathrm{ox}}
  - k_{\mathrm{mol}\rightarrow\mathrm{ox}} c_{\mathrm{mol}}.\label{eq:model-molecular}
\end{align}

Here, $c_{\mathrm{ox}}$ and $c_{\mathrm{mol}}$ are the tritium-atom concentrations assigned to the oxidised or fluoride-associated and molecular-hydrogen pools, respectively; $\bm{u}$ is the salt velocity; $D_{\mathrm{ox}}$ and $D_{\mathrm{mol}}$ are the corresponding effective diffusivities; and $f_{\mathrm{ox}}$ is the fraction of $S_T$ assigned to the oxidised pool after unresolved local relaxation. On a molar concentration basis, the corresponding SI units are $\mathrm{mol\,m^{-3}}$ for $c_{\mathrm{ox}}$ and $c_{\mathrm{mol}}$, $\mathrm{mol\,m^{-3}\,s^{-1}}$ for $S_T$, $\mathrm{m\,s^{-1}}$ for $\bm{u}$, $\mathrm{m^2\,s^{-1}}$ for $D_{\mathrm{ox}}$ and $D_{\mathrm{mol}}$, and $\mathrm{s^{-1}}$ for $k_{\mathrm{ox}\rightarrow\mathrm{mol}}$ and $k_{\mathrm{mol}\rightarrow\mathrm{ox}}$; $f_{\mathrm{ox}}$ is dimensionless.

The diffusion terms describe spatial transport within each pool using an effective Fickian relation, with the diffusive flux proportional to and directed down the corresponding concentration gradient. Interconversion between the pools is represented separately by the exchange terms. Where non-ideal thermodynamic effects can be represented by a scalar thermodynamic factor, that contribution should be included when assigning $D_{\mathrm{ox}}$ and $D_{\mathrm{mol}}$. It is not automatically included in an atomistic self-diffusivity. Significant coupling to independent gradients in salt composition or other species concentrations may require a more general flux relation. When $\bm{u}$ is a mean velocity in turbulent flow, unresolved turbulent mixing should also be represented in the pool fluxes, for example through an eddy diffusivity.

The limiting reference case $f_{\mathrm{ox}}=1$ assigns all newly produced tritium to the oxidised pool after local relaxation. Tritium can subsequently enter the molecular pool through the explicit interconversion terms. A non-zero molecular source fraction accounts for formation during unresolved local relaxation and does not imply immediate recombination of newly produced tritium atoms into \Ttwo{}. Sensitivity studies should vary $f_{\mathrm{ox}}$ over $0\leq f_{\mathrm{ox}}\leq1$ to assess the effect of the uncertain source partition.

Both concentrations use the same tritium-atom basis, so the exchange terms conserve total tritium. For a mixture of \HT{} and \Ttwo{}, $c_{\mathrm{mol}}=c_{\mathrm{HT}}+2c_{\mathrm{T_2}}$, with the individual concentrations expressed on a molecular molar basis. Solubility and partition relations must use the same basis. A single molecular-pool diffusivity and partition relation assume that \HT{} dominates or that the isotopologue mixture can be represented by effective coefficients.

The exchange terms use effective pseudo-first-order coefficients for a specified temperature and salt chemistry. Equilibrium between the pools is not imposed. An equilibrium approximation is justified when chemical relaxation is fast compared with transport, extraction and changes in local conditions. This reduction assumes dilute tritium and hydrogen-bearing reactant activities that are prescribed, approximately buffered or supplied by separate transport and chemistry balances. Significant hydrogen depletion or spatial variation in chemistry requires the coefficients to vary consistently or the relevant reactants to be resolved explicitly. At chemical equilibrium, they satisfy $k_{\mathrm{ox}\rightarrow\mathrm{mol}}c_{\mathrm{ox}}^{\mathrm{eq}}=k_{\mathrm{mol}\rightarrow\mathrm{ox}}c_{\mathrm{mol}}^{\mathrm{eq}}$. The equilibrium pool ratio therefore fixes the ratio of the two coefficients, but does not determine their absolute values.

If \TF{} and \Ttwo{} dominate the respective pools and both follow Henry's law, gas--salt equilibrium at fixed temperature and fluorine potential gives $c_{\mathrm{mol}}^{\mathrm{eq}}\propto(c_{\mathrm{ox}}^{\mathrm{eq}})^2$ \citep{Daramola2026FLiBe}. In this case, a constant ratio of exchange coefficients represents the equilibrium partition near a specified tritium loading. Modelling changes in the partition across a wider range of loadings requires a concentration-dependent equilibrium ratio.

For a closed, well-mixed system with constant coefficients and negligible source and boundary-transfer terms, the relaxation rate is $k_{\mathrm{ox}\rightarrow\mathrm{mol}}+k_{\mathrm{mol}\rightarrow\mathrm{ox}}$. In this limit, the equilibrium ratio and a species-resolved relaxation transient can in principle determine both coefficients. Experimental transients may also depend on diffusion, gas--liquid transfer and wall uptake or loss. These contributions need to be constrained independently or represented explicitly when fitting the exchange coefficients. Where hydrogen supply from the gas is not resolved, a fitted exchange coefficient may also include gas--liquid transfer resistance and need not be transferable between apparatuses. A resistance already included in that coefficient must not also be represented explicitly elsewhere in the model.

Atomistic calculations can help identify exchange mechanisms and estimate microscopic rates. Relating those rates to the two effective pool coefficients requires appropriate coarse-graining and validation.

At a stationary gas--liquid interface with no normal salt flow, let $\bm{n}$ denote the unit normal directed out of the salt. An effective gas-transfer condition for either pool is

\begin{equation}
  J_{i,g}=-D_i\nabla c_i\cdot\bm{n}
  =h_i(c_i-c_i^{*}),
  \qquad i\in\{\mathrm{ox},\mathrm{mol}\},
  \label{eq:gas-transfer}
\end{equation}

where $J_{i,g}$ is the outward tritium flux ($\mathrm{mol\,m^{-2}\,s^{-1}}$), $h_i$ is an effective mass-transfer coefficient on a liquid-concentration basis ($\mathrm{m\,s^{-1}}$), and $c_i^{*}$ is a salt-side reference concentration for the specified gas and salt conditions. The concentration $c_i$ is evaluated on the salt side of the resolved domain boundary. The coefficient $h_i$ represents the transfer resistance unresolved by the model. Where $c_i^{*}$ is an equilibrium concentration, the limit $h_i\to\infty$ gives $c_i=c_i^{*}$ at the boundary for finite flux. All concentrations and fluxes use a tritium-atom basis. For a molecular pool containing \HT{} and \Ttwo{} that obey separate Henry-law relations,

\begin{equation}
  c_{\mathrm{mol}}^{*}
  =K_{H,\mathrm{HT}}p_{\mathrm{HT}}
  +2K_{H,\mathrm{T_2}}p_{\mathrm{T_2}},
  \label{eq:gas-equilibrium}
\end{equation}

where the Henry coefficients are defined on the molecular molar basis used in \cref{eq:henry}. The factor of two accounts for the two tritium atoms in \Ttwo{}. The \Ttwo{} term in $c_{\mathrm{mol}}^{*}$ may be omitted when its effect on the boundary driving force is negligible. A single $h_{\mathrm{mol}}$ applies when the isotopologues share a transfer coefficient or a justified effective coefficient represents their combined flux. Equality of $c_{\mathrm{mol}}$ and $c_{\mathrm{mol}}^{*}$ can conceal opposing isotopologue fluxes. At equilibrium, both driving forces vanish. To calculate separate fluxes, obtain the individual salt-side concentrations and gas partial pressures from prescribed compositions, a speciation relation or separate balances. The total tritium-atom flux is then $J_{\mathrm{mol},g}=J_{\mathrm{HT},g}+2J_{\mathrm{T_2},g}$, with the individual fluxes on a molecular molar basis. A single $k_{L,\mathrm{mol}}a$ in \cref{eq:sparge-transfer} requires the same justification. The corresponding relation for the oxidised pool depends on salt and gas chemistry and need not follow Henry's law. An effective oxidised-pool flux can account for \TF{} recovered in the gas without assuming that molecular \TF{} is the dominant bulk transport species \citep{oishi1989tritiumRecovery,suzuki2000permeation}. Setting $h_{\mathrm{ox}}=0$ neglects direct release from this pool, although tritium can still be extracted following conversion to the molecular pool. The chemical form recovered in the gas may also reflect conversion at the interface.

When sparge-gas interfaces are not resolved geometrically, extraction can instead be represented by a volumetric term added to the corresponding pool balance:

\begin{equation}
  R_{i,\mathrm{sp}}
  =-k_{L,i}a(c_i-c_i^{*}),
  \qquad i\in\{\mathrm{ox},\mathrm{mol}\},
  \label{eq:sparge-transfer}
\end{equation}

where $k_{L,i}$ is an effective gas--liquid mass-transfer coefficient on a liquid-concentration basis ($\mathrm{m\,s^{-1}}$), $a$ is the gas--liquid interfacial area per unit salt volume ($\mathrm{m^{-1}}$), and $k_{L,i}a$ is the volumetric mass-transfer coefficient ($\mathrm{s^{-1}}$). The transfer term $R_{i,\mathrm{sp}}$ has units of $\mathrm{mol\,m^{-3}\,s^{-1}}$. The conventional notation $k_L$ for molecular extraction corresponds here to $k_{L,\mathrm{mol}}$. Setting $k_{L,\mathrm{ox}}=0$ restricts the unresolved sparging term to the molecular pool. Surface-flux and volumetric-transfer formulations may be used together for different interfaces, but should not double-count transfer across the same interfacial area. Bubble size and gas volume fraction affect the interfacial area, while bubble motion and the flow regime affect the mass-transfer coefficient. Transfer into bubbles should be distinguished from recovery in the extraction stream, which also depends on bubble residence time and gas--liquid separation \citep{evans2026sparging}.

At a stationary salt--wall interface with no normal salt flow, unresolved uptake or loss can be represented by

\begin{equation}
  J_{i,w}=-D_i\nabla c_i\cdot\bm{n}
  =h_{i,w}(c_i-c_i^{w,*}),
  \qquad i\in\{\mathrm{ox},\mathrm{mol}\},
  \label{eq:wall-transfer}
\end{equation}

where $c_i$ is evaluated on the salt side of the resolved domain boundary, $h_{i,w}$ is an effective transfer coefficient ($\mathrm{m\,s^{-1}}$) and $c_i^{w,*}$ is the salt-side concentration corresponding to the receiving wall state under the assumed partition relation.

Where the wall is not modelled explicitly, $h_{i,w}$ may combine species-dependent salt--wall reactions, transport through coatings and the wall, and receiving-side recombination or desorption. Trapping can influence this effective response, but a constant transfer coefficient does not resolve the associated storage and release history. Significant history dependence may require additional inventory variables or explicit trapping kinetics. When the wall or a coating is modelled explicitly, processes already represented in that domain must not also be included in $h_{i,w}$.

The no-flux condition is recovered when $h_{i,w}=0$. Negligible back-transfer may be approximated by $c_i^{w,*}=0$. When the structural material is modelled explicitly and interfacial storage is negligible, conservation of tritium requires

\begin{equation}
  J_{\mathrm{ox},w}+J_{\mathrm{mol},w}
  =-D_w\nabla c_w\cdot\bm{n},
  \label{eq:wall-flux-continuity}
\end{equation}

where $c_w$ is the mobile tritium-atom concentration in the wall, $D_w$ is its diffusivity, and the normal points from the salt into the wall. Fast equilibration along a single transfer pathway may justify an equilibrium partition relation, whose form depends on the salt and wall species. Where interfacial reactions limit transfer or competing pathways cannot be represented by a single partition relation, a reactive interface formulation is needed. Mixed isotopologues can also couple the interfacial H and T concentrations \citep{delaporteMathurin2026kineticInterface}.

\begin{figure}[htbp]
  \centering
  \includegraphics[width=\textwidth]{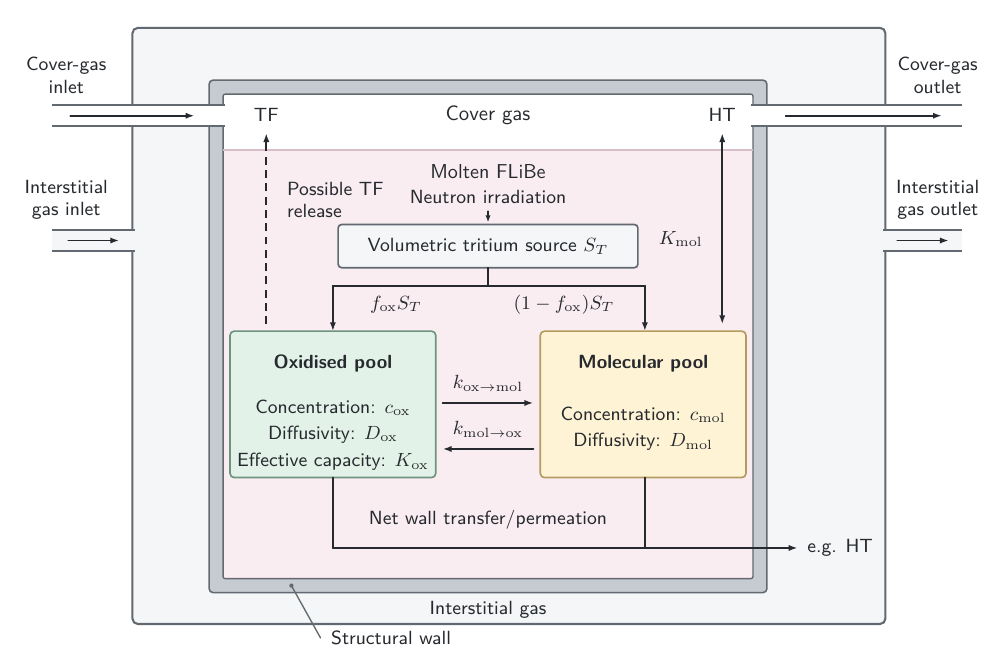}
  \caption{Schematic of the minimum two-pool continuum representation for tritium production, speciation, transport and loss in molten FLiBe. The volumetric source $S_T$ is partitioned between the oxidised or fluoride-associated pool and the molecular-hydrogen pool according to $f_{\mathrm{ox}}$, with effective interconversion coefficients $k_{\mathrm{ox}\rightarrow\mathrm{mol}}$ and $k_{\mathrm{mol}\rightarrow\mathrm{ox}}$. The pools have separate effective transport properties; $K_{\mathrm{ox}}$ denotes an effective capacity or partition relation for the oxidised pool under a specified salt chemistry, while $K_{\mathrm{mol}}$ denotes the molecular-pool gas--liquid equilibrium partition relation. The dashed arrow denotes possible TF release. Gas extraction and structural permeation are shown as boundary-transfer pathways; the wall arrow represents net tritium transfer, with HT shown as an example receiving-side gas product. The pools are effective continuum states and do not imply unique microscopic tritium species. The apparatus geometry is illustrative and is not intended to prescribe the neutron-source geometry; in blanket calculations, $S_T(\bm{x},t)$ is supplied by neutronics as a distributed volumetric source.}
  \label{fig:speciation-transport}
\end{figure}

Implementation requires the source and velocity fields, initial concentrations, pool diffusivities and exchange coefficients, together with the chosen transfer coefficients, partition relations and gas, wall, inlet and outlet conditions. Gas composition may be prescribed or determined from a gas-phase balance. For an \HT{}-dominated molecular pool obeying Henry's law, $K_{\mathrm{mol}}$ corresponds to $K_{H,\mathrm{HT}}$ under the same concentration and pressure convention. For the oxidised pool, $K_{\mathrm{ox}}$ denotes an effective capacity or partition relation for specified salt chemistry and gas environment. For a given local equilibrium state, the pool ratio implied by the exchange coefficients and the equilibrium partition relations must refer to the same temperature, redox condition, reactant activities and concentration basis. The source, exchange and boundary-transfer pathways are illustrated in \cref{fig:speciation-transport}.

Relevant open-source frameworks include FESTIM v2.0 for coupled hydrogen-isotope transport \citep{dark2026festimV2} and the MOOSE-based TMAP8 framework for tritium transport and fuel-cycle modelling \citep{simon2025tmap8}. In a recent preprint, Dark et al.\ coupled OpenMC, OpenFOAM and FESTIM to model neutron transport, thermal hydraulics and tritium transport in the Affordable, Robust, Compact (ARC) fusion-reactor concept \citep{dark2026arcPreprint}.

\subsection{Parameter extraction and uncertainty assessment}\label{subsec:parameter-table}

\Cref{tab:model-parameters} groups selected literature values by their role in the model and records their definitions, units, isotope or species basis, conditions, method and evidence class. Entries from different experiments or calculations are retained separately for comparison; they do not form a single transferable parameter set.

Uncertainty in these quantities arises from experimental measurement uncertainty, inference uncertainty introduced when effective coefficients are fitted or deconvolved, transferability uncertainty when a coefficient is applied beyond the salt, isotope, temperature or boundary conditions under which it was obtained, and model-form uncertainty associated with unresolved species or transport pathways. These contributions should be distinguished where possible. Numerical uncertainties should be reported with their type and any stated confidence or coverage level; missing uncertainty estimates should be identified. When one of the quantities in $P_{\mathrm{liq}}=DK$ is derived from the other two, the three must not be sampled independently. Reported covariance between fitted parameters, including Arrhenius prefactors and activation energies, should also be retained. Where covariance information is unavailable, source-specific parameter sets can be examined as alternative scenarios, but the spread between them is not a statistical confidence interval. Parameter identifiability should also be assessed when several effective coefficients are inferred from the same integrated transient: agreement with a single release or permeation history does not necessarily identify bulk diffusion, pool interconversion, gas--liquid transfer and wall-transfer resistances independently.

For compactness, Arrhenius correlations in \cref{tab:model-parameters} use $X(T)=X_0\exp[-E/(RT)]$, with $T$ in kelvin and $R$ the molar gas constant. A reported liquid permeability $P_{\mathrm{liq}}=DK$ is retained as a combined transport observable rather than an independent input to \cref{eq:model-oxidised,eq:model-molecular}. Concentration, isotope and molecular or atomic conventions are stated explicitly because nominally similar coefficients are not necessarily comparable.

\begingroup
\footnotesize
\renewcommand{\arraystretch}{1.16}
\setlength{\tabcolsep}{2.5pt}
\begin{longtable}{@{}P{0.105\textwidth}P{0.080\textwidth}P{0.115\textwidth}P{0.225\textwidth}P{0.12\textwidth}P{0.265\textwidth}@{}}
\caption{Selected source-specific quantities grouped by their role in the two-pool continuum model of tritium in molten FLiBe. Values are retained under their reported definitions and are not, by themselves, recommended model inputs. Assignment of an inferred quantity to a pool retains the source's species interpretation and is conditional on its experimental and modelling assumptions; the introduced gas does not independently establish the dissolved transport species.}\label{tab:model-parameters}\\
\toprule
\textbf{Quantity} &
\textbf{Symbol} &
\makecell[l]{\textbf{Pool or}\\\textbf{species}} &
\makecell[l]{\textbf{Value / relation}\\\textbf{conditions}} &
\makecell[c]{\textbf{Evidence}\\\textbf{class}} &
\makecell[l]{\textbf{Applicability}\\\textbf{and limitations}} \\
\midrule
\endfirsthead
\toprule
\textbf{Quantity} &
\textbf{Symbol} &
\makecell[l]{\textbf{Pool or}\\\textbf{species}} &
\makecell[l]{\textbf{Value / relation}\\\textbf{conditions}} &
\makecell[c]{\textbf{Evidence}\\\textbf{class}} &
\makecell[l]{\textbf{Applicability}\\\textbf{and limitations}} \\
\midrule
\endhead

\multicolumn{6}{@{}l}{\textit{Source and model-closure inputs}}\\*
\addlinespace[2pt]

Volumetric source &
$S_T$ &
Total tritium &
System-specific; calculate for the specified neutron spectrum, flux, geometry and lithium-isotope composition &
Calculated or derived &
Use the same tritium-atom basis as $c_{\mathrm{ox}}$ and $c_{\mathrm{mol}}$. Distinguish local production rate from a volume-averaged source and retain the irradiation history. See \citet{sawan2003transmutation,dolan2021tritium,ferry2023libra} for examples of tritium production calculations. \\

Initial source partition &
$f_{\mathrm{ox}}$ &
Oxidised and molecular-hydrogen pools &
Limiting reference case $f_{\mathrm{ox}}=1$; vary $0\leq f_{\mathrm{ox}}\leq1$ in sensitivity studies &
Assumed model input &
The post-relaxation source partition has not been measured directly. Its value may depend on salt composition, redox condition and the chemical relaxation of newly produced tritium. Constraining this partition requires a description of newly produced tritium and a defined time over which its local relaxation is included in the source term. Equilibrium speciation calculations alone are insufficient. \\

\addlinespace[3pt]
\multicolumn{6}{@{}l}{\textit{Molecular-pool properties}}\\*
\addlinespace[2pt]

Molecular solubility &
$K_{\mathrm{mol}}$ &
\ce{H2}/\ce{D2} analogue &
Combined \ce{H2}/\ce{D2} correlation in molten \ce{Li2BeF4}, 773--973 K. At 773, 873 and 973 K, the correlation gives $10^3K_c=1.36$, 2.60 and 4.39, respectively \citep{malinauskas1974solubilities}. &
Measured uptake; fitted correlation &
$K_c=c_{\mathrm{liquid}}/c_{\mathrm{gas}}$ is dimensionless. The values shown are calculated from the combined correlation, rather than individual measurements. Provides molecular-H analogue evidence; retain the isotope identity and source convention. \\

Molecular solubility &
$K_{\mathrm{mol}}$ &
\ce{H2} analogue &
$K_0=6.57\times10^{2}$ \si{\mole\per\cubic\metre\per\pascal}; $E=104.8$ kJ mol$^{-1}$; 773--873 K \citep{nakamura2015hydrogen} &
Fitted or inferred &
Derived from permeability and diffusivity for FLiBe (reported LiF:BeF$_2$ ratio 67.2:32.8) purified by bubbling with a gas containing 1000 ppm HF. The steady-state flux relation uses the difference between the upstream and downstream \ce{H2} partial pressures. The authors inferred molecular transport from the approximately linear dependence of the permeation flux on that pressure. Nakamura et al.\ use both $j_{\mathrm{H}}$ and $j_{\mathrm{H_2}}$ for the steady-state flux, leaving it unclear whether moles refer to H atoms or \ce{H2} molecules \citep{saraswat2026permeation}. \\

Molecular diffusivity &
$D_{\mathrm{mol}}$ &
\ce{H2} analogue &
$D_0=2.09\times10^{-8}$ \si{\square\metre\per\second}; $E=25.2$ kJ mol$^{-1}$; 773--873 K \citep{nakamura2015hydrogen} &
Fitted or inferred &
Obtained from transient permeation in a static cylindrical FLiBe apparatus. Treat separately from atomistic self-diffusion and from diffusivities inferred using different boundary assumptions. \\

Molecular diffusivity &
$D_{\mathrm{mol}}$ &
Neutral ${}^{3}\mathrm{H}^{0}$ (molecular \ce{H2} state) &
At 973, 1173 and 1373 K: $6.3\pm1.4$, $16.3\pm3.3$ and $27.0\pm4.1$, respectively, in units of $10^{-9}$ \si{\square\metre\per\second}; $E_D=40.0\pm3.0$ kJ mol$^{-1}$ \citep{lam2021hydrogenValence} &
AIMD prediction &
AIMD self-diffusion in clean FLiBe for the imposed neutral state; the neutral state exists as \ce{H2} in the simulation. Do not equate directly with an experimental effective diffusivity. \\

Liquid permeability &
$P_{\mathrm{liq}}$ &
\ce{H2} analogue &
$P_0=1.37\times10^{-5}$ \si{\mole\per\metre\per\second\per\pascal}; $E=129.7$ kJ mol$^{-1}$; 773--873 K \citep{nakamura2015hydrogen} &
Fitted or inferred &
Determined from steady-state permeation measurements, with $P_{\mathrm{liq}}=D K$. The molar basis is ambiguous for the same reason as in the Nakamura solubility entry. Retain as a combined transport observable rather than an independent continuum-model input. \\

Liquid permeability &
$P_{\mathrm{liq}}$ &
\ce{H2}/\ce{D2} analogue &
Salt-side charging: \ce{H2}, 773--873 K; \ce{D2}, 773--973 K. At 873 K, the reported \ce{H2} fits give $P_{\mathrm{liq}}=7.72\times10^{-14}$ and $8.45\times10^{-14}$ \si{\mole\per\metre\per\second\per\pascal} for coating permeation-reduction factors of $\infty$ and 1, respectively \citep{saraswat2026permeation}. &
Fitted or inferred &
The displayed values are calculated from the source correlations and converted from molecular counts to moles of \ce{H2}. Reported combined permeability uncertainty is within $\pm41\%$; the two coating cases are alternative boundary assumptions, not confidence limits. Salt wetting, coverage and bubbles affect inference; metal-side charging reduced apparent permeability by up to 77\% relative to salt-side estimates. \\

\addlinespace[3pt]
\multicolumn{6}{@{}l}{\textit{Oxidised or fluoride-associated pool properties}}\\*
\addlinespace[2pt]

Oxidised capacity &
$K_{\mathrm{ox}}$ &
HF/DF analogue &
66:34 mol\% \ce{LiF}--\ce{BeF2}, 500--700\,\si{\celsius}. At 600\,\si{\celsius}, the HF and DF coefficients are $(2.16\pm0.05)\times10^{-4}$ and $(1.83\pm0.03)\times10^{-4}$, respectively, in moles of solute per mole of melt per atmosphere \citep{field1967solubilities}. &
Measured uptake; fitted solubility &
Reported uncertainties are 95\% confidence limits on the fitted values. Retain the original solute/melt molar basis; conversion to a volumetric coefficient requires a consistent melt molar density. These analogue coefficients are not a universal effective $K_{\mathrm{ox}}$. \\

Effective oxidised capacity &
$K_{\mathrm{ox}}$ &
D, fluoride-associated interpretation &
873 K: $3.1\times10^{-4}$ \si{\mole\per\cubic\metre\per\pascal}; 923 K: $1.0\times10^{-4}$ \si{\mole\per\cubic\metre\per\pascal} \citep{anderl2004deuterium} &
Fitted or inferred &
The fit uses the upstream \ce{D2} partial pressure in the nickel source probe. The fitted diffusivities and solubilities led the authors to infer fluoride-associated transport. The Field--Shaffer \ce{DF} coefficient refers to \ce{DF} partial pressure at gas--salt equilibrium. \\

Oxidised diffusivity &
$D_{\mathrm{ox}}$ &
D, fluoride-associated interpretation &
873 K: $8.0\times10^{-10}$ \si{\square\metre\per\second}; 923 K: $3.0\times10^{-9}$ \si{\square\metre\per\second} \citep{anderl2004deuterium} &
Fitted or inferred &
JUPITER-II effective diffusivities from an analytical fit to cylindrical permeation data. Retain the authors' species interpretation and apparatus assumptions. \\

Oxidised diffusivity &
$D_{\mathrm{ox}}$ &
Protonic \ce{^3H+} &
At 973, 1173 and 1373 K: $3.0\pm0.5$, $7.0\pm1.3$ and $11.1\pm1.6$, respectively, in units of $10^{-9}$ \si{\square\metre\per\second}; $E_D=36.0\pm2.0$ kJ mol$^{-1}$ \citep{lam2021hydrogenValence} &
AIMD prediction &
AIMD self-diffusion for the imposed protonic state; transport is coupled to fluoride coordination. Method-, composition- and charge-state-specific. \\

\addlinespace[3pt]
\multicolumn{6}{@{}l}{\textit{Direct-tritium integrated transport}}\\*
\addlinespace[2pt]

Integrated tritium transport &
\makecell[c]{$P_{\mathrm{liq}}$\\$D,\ K$} &
Direct T; microscopic pool unresolved &
Nominal 2:1 FLiBe. Reported fits: $D_0=9.3\times10^{-7}$ \si{\square\metre\per\second}, $E_D=42$ kJ mol$^{-1}$; $K_0=7.9\times10^{-2}$ \si{\mole\per\cubic\metre\per\pascal} as printed, $E_K=35$ kJ mol$^{-1}$. Diffusivity measurements cover 823--973 K; the abstract gives 773--973 K for the experiment \citep{calderoni2008permeation}. &
Fitted or inferred &
\ce{T2} was supplied below the nickel membrane, with 5 vol\% \ce{H2} in the sweep gas. The fit uses the square root of the upstream \ce{T2} partial pressure. On a molar-flux basis, the source equation requires $K$ in $\mathrm{mol\,m^{-3}\,Pa^{-1/2}}$. The numerical fit is retained as reported and is not treated as a Henry-law coefficient. The fits assume one-dimensional diffusion and negligible resistance of the nickel membrane; statistical uncertainties for the fitted parameters are not reported. The coefficients cannot be assigned uniquely to either model pool. \\

\addlinespace[3pt]
\multicolumn{6}{@{}l}{\textit{Exchange and interfacial transfer}}\\*
\addlinespace[2pt]

Forward exchange &
$k_{\mathrm{ox}\rightarrow\mathrm{mol}}$ &
Oxidised to molecular-hydrogen &
Fit an effective coefficient for the specified hydrogen and salt conditions; do not assume an intrinsic transferable rate \citep{terai1996intrex,suzuki1998speciesChange} &
Fitted or inferred &
Hydrogen addition strongly changes HT release, but reported transients combine chemical exchange, gas--liquid transfer, mixing and bulk transport. Use a pseudo-first-order coefficient only for the conditions from which it was inferred. \\

Gas--liquid transfer &
$h_{\mathrm{mol}}$ &
HT under \ce{H2} purge or bubbling &
Overall coefficient $(0.75\text{--}1.1)\times10^{-5}\ \mathrm{m\,s^{-1}}$ at 873 K \citep{suzuki1998htRelease} &
Fitted or inferred &
Obtained from in-pile HT release from approximately 200 g FLiBe and varied with \ce{H2} bubbling rate. The authors attributed the limiting process to HT transport through the FLiBe liquid film. Use this apparatus-specific overall coefficient with the bulk-liquid concentration when the liquid film is unresolved and the bubbling conditions are comparable. \\

\bottomrule
\end{longtable}
\endgroup

At 873 K, the Malinauskas--Richardson \ce{H2}/\ce{D2} Ostwald correlation gives $K_H=K_c/(RT)=3.58\times10^{-7}\ \mathrm{mol\,m^{-3}\,Pa^{-1}}$ for an ideal gas at the salt temperature \citep{malinauskas1974solubilities}. In the same nominal units, the permeation fits give $3.52\times10^{-4}$ for Nakamura et al.\ and $3.1\times10^{-4}$ for Anderl et al. Both are about three orders of magnitude larger than the Ostwald-based estimate. The authors inferred different dissolved carriers despite the similar magnitudes of the two fitted coefficients \citep{nakamura2015hydrogen,anderl2004deuterium}.

Between 873 and 923 K, the fitted $D$ from Anderl et al.\ increases and the fitted $K$ decreases by factors greater than three. Their product rises by about 20\%, from $2.5\times10^{-13}$ to $3.0\times10^{-13}\ \mathrm{mol\,m^{-1}\,s^{-1}\,Pa^{-1}}$. Use the fitted $D$ and $K$ as pairs at their reported temperatures \citep{anderl2004deuterium}.

Combine published coefficients only when they describe the same physical quantity on a common species and isotope basis and apply to the salt chemistry, temperature and transport regime being modelled. Where several source-specific relations are plausible, evaluate each separately and report how the choice affects predicted tritium inventory, release and wall losses.

\section{Validation gaps and experimental priorities}\label{sec:validation}

Existing datasets are difficult to combine because salt preparation, tritium introduction route, redox condition, gas environment, hydrodynamics and boundary materials vary between experiments \citep{terai1996intrex,anderl2004deuterium,dolan2023experimental,saraswat2026permeation}. We therefore separate validation into controlled property measurements, coupled cells and interfacial experiments, flowing-salt loops, and neutron-irradiated integrated tests \citep{ferry2023libra,gilbert2026librti,fuerst2025mstte,delaporteMathurin2025baby,delaporteMathurin2026baby1l}.

Controlled cells can constrain molecular and oxidised solubility, diffusion and speciation under specified salt chemistry. Bulk-property and interfacial-transfer measurements are distinguished here by the quantities they target. Both involve interfaces, whose effects must be accounted for when inferring bulk properties. Microscopic simulations are most informative when confronted directly with structure- or dynamics-sensitive experimental observables and should not be treated as standalone validation evidence \citep{armstrong2020dynamic,fayfar2024complex}. Gas--liquid exchange and wall-transfer experiments can then calibrate or cross-check interfacial descriptions while isolating them as far as possible from bulk transport. Flowing loops extend this validation to advection, mixing, extraction and distributed wall losses, while neutron-irradiated integrated experiments test the combined source, chemical relaxation, transport, extraction and accountancy description at system level. Wherever possible, the dataset used to infer a parameter should be distinguished from the independent dataset used to validate the resulting model.

Computational work should target the quantities that remain poorly constrained, such as the relative stability and chemical potentials of tritium-bearing states, their diffusivities and exchange mechanisms, and their interactions with gas and structural surfaces. Calculations need to cover the salt compositions and redox conditions relevant to the experiments. First-principles methods and validated interatomic potentials can help where measurements are scarce, but microscopic predictions still need to be related to the effective parameters used in the continuum model. Coupled transport calculations, sensitivity analysis and uncertainty quantification can then identify which uncertainties have the greatest effect on predicted tritium inventory, release and extraction.

Calculations for molten NaCl have combined molecular sampling, neural-network potentials and higher-level electronic-structure corrections with quasichemical theory to predict excess chemical potentials \citep{shi2022moltenThermodynamics}. A recent preprint also applies hybrid quantum--classical calculations to tritium binding in finite FLiBe clusters. The quantum-assisted results closely reproduce their embedded classical references, but substantial fragmentation errors remain and the study does not yet provide bulk speciation free energies \citep{das2026quantum}.

The Molten Salt Tritium Transport Experiment (MSTTE) is a forced-convection fluoride-salt loop at Idaho National Laboratory designed for hydrogen-isotope transport experiments, with an initial phase intended to use FLiNaK (\ce{LiF}--\ce{NaF}--\ce{KF}) and deuterium as surrogates. The reported design provides for deuterium injection into flowing FLiNaK and for measurements of permeation through structural materials and evolution from free surfaces, offering a prospective benchmark for advection, hydrodynamics and coupled interfacial transport \citep{fuerst2025mstte,bowers2026mstteCFD}. Its relevance here is therefore apparatus-level validation rather than direct measurement of FLiBe tritium properties.

Earlier INTREX-FLiBe and JUPITER-II campaigns established much of the experimental basis for FLiBe tritium chemistry and transport. Neutron-irradiation studies demonstrated the sensitivity of recovered chemical form and release to purge-gas composition and redox condition, while JUPITER-II permeation experiments coupled transport through the salt to metallic barriers and inferred transport quantities from integrated transients \citep{terai1996intrex,suzuki1998htRelease,suzuki1998speciesChange,suzuki1998mechanism,terai2001release,fukada2002initial,fukada2003flibeD2,anderl2004deuterium,petti2006jupiter,calderoni2008permeation,abe2008jupiter}. These datasets remain important for mechanism discrimination, but their fitted coefficients should be interpreted with the associated species and apparatus assumptions intact.

The MIT Reactor irradiation programme provides an important intermediate benchmark between property measurements and integrated fusion experiments. In-core irradiations of \ce{^{7}Li}-enriched FLiBe compared calculated tritium generation with measured release and demonstrated that retention in graphite can represent a significant part of the inventory \citep{dolan2020graphite,dolan2021tritium}. A subsequent neutron-irradiated FLiBe experiment measured both time-dependent free-surface release and permeation through the stainless-steel test assembly and compared these observations with a coupled multiphysics transport model \citep{dolan2023experimental}. An independent estimate of tritium production helps interpret measurements of release, retention and wall permeation in terms of the tritium balance.

The Liquid Immersion Blanket: Robust Accountancy (LIBRA) project provides a route towards integrated FLiBe breeding and accountancy tests. The published BABY 100 mL and BABY 1L campaigns used ClLiF and therefore do not provide direct FLiBe material-property measurements, but they provide integrated benchmarks for breeding estimates, time-resolved recovery and gas handling under \SI{14}{\mega\electronvolt} D--T neutron irradiation \citep{delaporteMathurin2025baby,delaporteMathurin2026baby1l}. BABY 1L targeted a salt temperature of 700\,\si{\celsius}, but the minimum temperature in the salt at steady state was approximately 630\,\si{\celsius}. Release under inert gas was consistent with diffusion limitation, and hydrogen addition accelerated recovery \citep{delaporteMathurin2026baby1l}. Transferring results from such experiments to blanket conditions nevertheless requires care because geometry, surface-area-to-volume ratio, flow regime, wall material and residence time can change the balance of transport resistances. Bulk diffusivities or equilibrium partition coefficients may be transferable when species and salt chemistry are comparable, whereas $k_L a$, fitted release constants and effective wall-transfer coefficients are generally apparatus-specific \citep{saraswat2026permeation,park2026crossScale,yang2026multidimensional}. Recent high-fidelity FLiBe contactor modelling found a factor-of-12 variation in predicted normalised extraction rate across the solubility correlations examined, illustrating how property uncertainty can propagate to extraction predictions \citep{li2026gasLiquidExtraction}.

A complementary approach is to validate predictions directly against extraction-system performance. Measurements of tritium extraction efficiency across controlled variations in temperature, salt composition, inlet tritium concentration, salt and gas flow rates, and contactor geometry can constrain system behaviour even when individual transport parameters remain uncertain. Tritium recovered in the extraction stream should be distinguished from the total amount leaving the salt, which may also include wall uptake and other losses. Empirical correlations derived from these measurements can be compared with mechanistic predictions and tested against independent operating conditions, although extrapolation to other geometries or salt conditions requires further validation.

\begin{figure}[htbp]
  \centering
  \includegraphics[width=\textwidth,trim=0pt 0 1pt 0,clip]{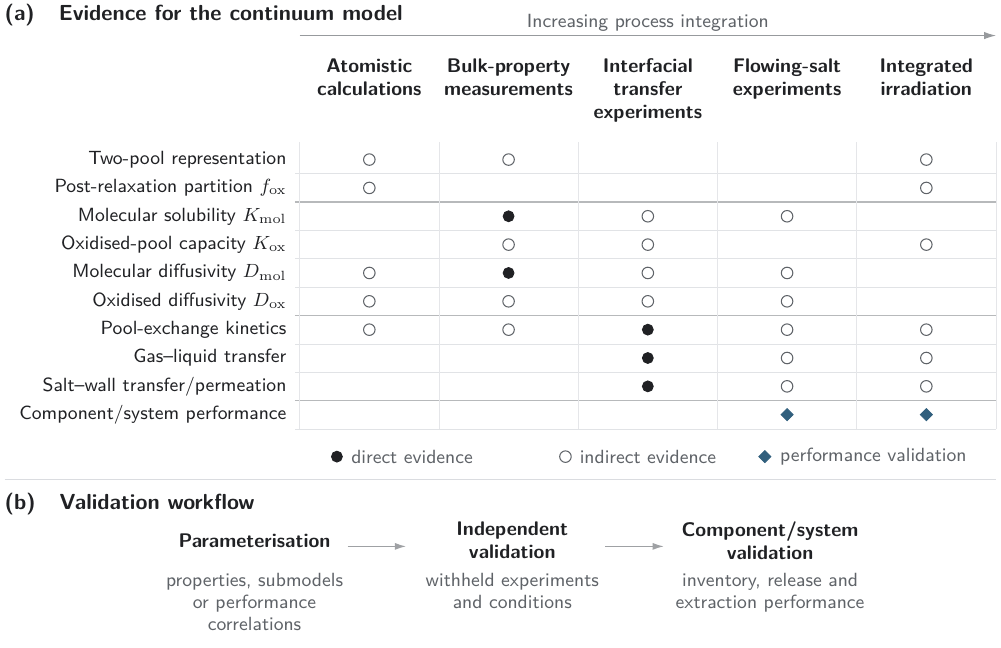}
  \caption{Sources of evidence for the two-pool continuum model of tritium transport in molten FLiBe. \textbf{(a)} Rows are grouped into model structure and source partition, pool properties, exchange and boundary transfer, and component or system performance. Columns identify complementary methods and do not prescribe an experimental sequence. Integrated irradiation tests combine neutron-driven tritium production with chemical relaxation, transport, release and retention. Bulk-property and interfacial-transfer measurements are distinguished by their target quantities; both involve interfaces. Filled circles identify methods that can provide direct evidence for a model quantity or assumption, open circles denote indirect evidence, and diamonds denote potential validation of component or system performance against independent measurements. Blank cells indicate no assigned evidential role. The symbols describe potential roles under suitably controlled conditions and do not assess the availability or quality of existing datasets. \textbf{(b)} Illustrative workflow distinguishing parameterisation, independent validation against withheld experiments and conditions, and assessment of component or system performance. Parameterisation may use property measurements, submodel data or prototype performance correlations.}
  \label{fig:validation-framework}
\end{figure}

\Cref{fig:validation-framework} links the model quantities to potential sources of evidence and summarises the validation workflow. \Cref{tab:experimental-comparison} compares the selected experimental platforms.

\begingroup
\footnotesize
\renewcommand{\arraystretch}{1.15}
\setlength{\tabcolsep}{3pt}

\begin{longtable}{@{}P{0.16\textwidth}P{0.18\textwidth}P{0.20\textwidth}P{0.20\textwidth}P{0.20\textwidth}@{}}

\caption{Comparison of selected experimental platforms and the evidence they can provide for tritium-transport models in molten FLiBe.}
\label{tab:experimental-comparison}\\

\toprule
\textbf{Experiment} &
\textbf{Salt / isotope source} &
\textbf{Conditions and materials} &
\textbf{Main observables} &
\textbf{Evidence for the model} \\
\midrule
\endfirsthead

\toprule
\textbf{Experiment} &
\textbf{Salt / isotope source} &
\textbf{Conditions and materials} &
\textbf{Main observables} &
\textbf{Evidence for the model} \\
\midrule
\endhead

\multicolumn{5}{@{}l}{\textit{Experiments using FLiBe}}\\
\addlinespace[2pt]

INTREX-FLiBe &
$\sim$100 g 2:1 FLiBe; neutron-generated T &
YAYOI in-pile irradiation; 873 K; purge gas flowed over the molten salt &
Time-resolved tritium release; HT/TF fractions; response to purge-gas composition and high-temperature dehumidification time &
Constrains chemical-form and release response, but not a unique bulk $D$ or $K$ \citep{terai1996intrex} \\

\addlinespace[2pt]

JUPITER-II D &
$\sim$400 cm$^3$ 2:1 FLiBe; D analogue &
Type 316 stainless-steel pot, nickel crucible and nickel permeation probes; 600 and 650\,\si{\celsius}; Ar purge with quadrupole mass spectrometry (QMS) analysis &
D permeation transients; fitted diffusivity and solubility coefficients at both temperatures &
Effective $D$ and $K$ from the nickel-fed permeation fit, with fluoride association inferred \citep{anderl2004deuterium} \\

\addlinespace[2pt]

JUPITER-II T &
Nominal 2:1 FLiBe; controlled \ce{T2} source &
550--700\,\si{\celsius}; stainless-steel assembly with nickel salt-contact surfaces and a 2 mm Nickel 200 membrane; initial FLiBe layer thickness 8.1 mm, increased to 13.7 mm for selected low-pressure measurements; Ar/\ce{T2} source gas; Ar with 5 vol\% \ce{H2} sweep gas; QMS, gas chromatography and proportional counting &
Steady-state tritium permeability; transient diffusivity; derived solubility &
Direct-T integrated transport. Constrains the combined transport description but does not uniquely separate oxidised and molecular pools \citep{calderoni2008permeation} \\

\addlinespace[2pt]

MITR in-core FS-1--3 &
Regenerated \ce{^{7}Li}-enriched FLiBe from the Molten Salt Reactor Experiment secondary coolant; neutron-generated T &
MITR in-core irradiations at 700\,\si{\celsius}; material coupons included to examine tritium transport and retention &
Calculated tritium production compared with measured release. Retained tritium assessed by post-irradiation thermal desorption of graphite &
Integrated production, release and retention benchmark; constrains material sinks and accountancy rather than an isolated FLiBe transport coefficient \citep{carpenter2017partitioning,dolan2020graphite,dolan2021tritium} \\

\addlinespace[2pt]

MITR ex-core transport &
2:1 FLiBe; neutron-generated T &
Ex-core neutron beam-port irradiation of molten FLiBe in a stainless-steel test stand &
Time-dependent free-surface release and wall permeation, compared with a coupled multiphysics transport model &
Benchmark for models coupling bulk transport, free-surface release and wall permeation \citep{dolan2023experimental} \\

\addlinespace[2pt]

HYPERION &
2:1 FLiBe; \ce{H2}/\ce{D2} analogue &
773--973 K. Nickel 200 permeation cell with an external thermally sprayed \ce{Al2O3} coating and salt-side and metal-side charging configurations. After stirring, the final 49.9 g FLiBe loading formed a layer about 5 mm thick that fully covered the membrane &
Steady-state permeation; isotope dependence; sensitivity to salt wetting, membrane coverage and interfacial gas bubbles &
Constrains the integrated H/D permeation response and its sensitivity to boundary conditions; H/D analogue rather than a direct tritium measurement \citep{saraswat2026permeation} \\

\midrule
\multicolumn{5}{@{}l}{\textit{Flowing, integrated and complementary platforms}}\\*[2pt]

MSTTE (initial FLiNaK/D phase) &
FLiNaK; D surrogate &
Forced-convection type 316 stainless-steel loop designed for deuterium injection and intended for metal-permeation/free-surface measurements &
Target measurements of deuterium permeation through metal and release at the free surface &
Prospective benchmark for advection, wall permeation and free-surface release in flowing FLiNaK \citep{fuerst2025mstte,bowers2026mstteCFD} \\

\addlinespace[2pt]

LIBRA (design basis) &
Planned $\sim$1000 kg FLiBe; D--T neutron source &
Integrated large-volume FLiBe concept with extraction, redox/radiochemistry control and tritium accountancy &
Planned tritium breeding estimates and measurements of recovery and chemical form for system-wide tritium accountancy &
Future system-level FLiBe model validation; the cited publication establishes the experimental concept and objectives rather than a completed property dataset \citep{ferry2023libra} \\

\addlinespace[2pt]

BABY \SI{100}{mL} &
190 g ClLiF; neutron-generated T &
30.5 mol\% LiF--69.5 mol\% LiCl; 700\,\si{\celsius}; Inconel 625 crucible; D--T neutron irradiation; He sweep &
Time-resolved tritium recovery, with the total gas-phase yield used to estimate the tritium breeding ratio (TBR). In the reported runs, all collected tritium was water-insoluble, consistent with \HT{} or \Ttwo{} &
ClLiF surrogate-salt benchmark for recovery and accountancy. The recovered inventory fell below calculated tritium production \citep{delaporteMathurin2025baby} \\

\addlinespace[2pt]

BABY 1L &
1.88 kg ClLiF; neutron-generated T &
30.5 mol\% LiF--69.5 mol\% LiCl. The target salt temperature was 700\,\si{\celsius}. The minimum temperature in the salt at steady state was about 630\,\si{\celsius}. Four D--T irradiation runs covered inert and hydrogen-containing carrier-gas conditions &
TBR estimated from recovered tritium and compared with OpenMC. Release kinetics, chemical form of recovered tritium and response to hydrogen addition &
Surrogate-salt benchmark for breeding estimates, time-resolved recovery from the inner and outer vessels, and recovered chemical form. Release under inert gas was consistent with diffusion limitation. The authors proposed isotope exchange with tritium retained on apparatus surfaces to explain the additional outer-vessel release after hydrogen addition \citep{delaporteMathurin2026baby1l} \\

\bottomrule
\end{longtable}

\endgroup

The measurements that would most directly reduce uncertainty in the continuum model are summarised in \cref{tab:priority-measurements}.

\begin{table}[htbp]
\centering
\caption{Priority measurements for reducing uncertainty in the two-pool continuum model.}
\label{tab:priority-measurements}
\footnotesize
\renewcommand{\arraystretch}{1.18}
\setlength{\tabcolsep}{4pt}
\begin{tabularx}{\textwidth}{@{}P{0.27\textwidth}P{0.43\textwidth}Y@{}}
\toprule
\textbf{Priority measurement} &
\textbf{Required controls and observables} &
\textbf{Model quantities informed} \\
\midrule

Salt redox state or fluorine potential &
Record preparation and purification history, impurities, temperature, gas composition and redox condition alongside the transport measurement. For electrochemical measurements, specify the reference electrode or redox couple, calibration and relevant activity convention &
Pool-state interpretation, exchange behaviour and parameter transferability \\

\addlinespace[3pt]

Molecular-hydrogen solubility and diffusion &
Measure equilibrium uptake and transient transport separately for identified hydrogen isotopologues in characterised FLiBe over a defined temperature range &
$K_{\mathrm{mol}}(T)$ and $D_{\mathrm{mol}}(T)$ \\

\addlinespace[3pt]

Non-molecular tritium uptake, mobility and speciation &
Measure retained inventory and time-resolved transport under controlled redox and temperature conditions. Use species-sensitive measurements to test for oxidised and reduced states &
Bounds on $K_{\mathrm{ox}}(T)$ and $D_{\mathrm{ox}}(T)$ where the oxidised pool dominates, and adequacy of the two-pool model \\

\addlinespace[3pt]

Oxidised--molecular exchange kinetics &
In a well-mixed cell with negligible tritium production and boundary transfer, measure the equilibrium pool ratio and the relaxation of the pool concentrations after a perturbation. Hold temperature and reactant activities fixed during relaxation. For open-cell experiments, constrain gas--liquid transfer and wall uptake independently when fitting the exchange rates &
$k_{\mathrm{ox}\rightarrow\mathrm{mol}}$ and $k_{\mathrm{mol}\rightarrow\mathrm{ox}}$ \\

\addlinespace[3pt]

Gas--liquid mass transfer &
Control interfacial area, gas composition and flow, mixing and temperature. Measure time-resolved transfer separately from wall loss and determine the salt-side driving force $c_i-c_i^{*}$ &
Volumetric transfer coefficient $k_{L,i}a$ or surface-transfer coefficient $h_i$ \\

\addlinespace[3pt]

Salt--wall transfer and permeation &
Measure receiving-side flux over a range of upstream hydrogen-isotope partial pressures. Characterise the salt-side and receiving-side conditions, including salt redox state. Record wall material, thickness and surface state. Use tests at several wall thicknesses or independent measurements to help distinguish the resistances &
Apparent pressure dependence and effective wall-transfer coefficients $h_{i,w}$. Wall permeability and interfacial partition or kinetics where separately constrained \\

\addlinespace[3pt]

Integrated irradiation experiment &
Constrain tritium production independently of the release measurements. Measure time-resolved release, retained inventory, the chemical form of released tritium and wall loss. Record the temperature and flow histories &
Joint constraints on source partition, exchange and boundary-transfer descriptions; overall model-form adequacy \\

\bottomrule
\end{tabularx}
\end{table}

\section{Conclusions}\label{sec:conclusions}

Separating tritium into an oxidised or fluoride-associated pool and a molecular-hydrogen pool provides a practical continuum description where states omitted from these pools do not materially affect inventory or transport. The oxidised pool need not correspond to one microscopic species under all conditions. Reduced or Be--H-associated states under strongly reducing conditions remain an additional model-form uncertainty and are not separately parameterised at present. \HT{} is the best-supported molecular extraction species when protium is present, while the identity and transport mechanism of non-molecular tritium remain less certain.

Published solubility, diffusion, release and permeation data cannot be combined consistently without tracking the introduced isotope-bearing species, salt composition and preparation, redox condition, gas environment, hydrodynamics and boundary materials. In particular, atomistic self-diffusion, fitted release diffusivity and integrated permeation are different quantities; comparisons require consistent definitions and a transport model that accounts for the processes represented by each.  

The parameters of this two-pool advection--diffusion--reaction model, or its reaction--diffusion limit for static salt, should be reported with their evidential basis and uncertainty. Independent validation should test whether the chosen pool structure and transport closures remain adequate across the chemical and operating conditions of interest.

Future integrated breeder experiments should constrain tritium production independently and measure release, retention and wall losses. Measurements of redox state, species conversion and gas--liquid transfer are needed to distinguish competing transport and speciation models.

\section*{Competing interests}

The authors declare no competing interests.

\section*{Acknowledgements}

This work was supported by the EPSRC Fusion Grant 2022/27 [EP/W006839/1]. This work is part of the UKAEA's Lithium Breeding Tritium Innovation (LIBRTI) programme, funded by the Department for Energy Security and Net Zero, as announced by the UK Government in October 2023. Microsoft Copilot and OpenAI Codex were used for reference checking and editing, Python code refinement and LaTeX formatting. The authors take full responsibility for the final manuscript.

\section*{Data availability}

No new experimental or simulation data were generated for this review. All data supporting the analysis are included within the article or available from the cited sources. Further information on the compilation of the figures and tables is available upon reasonable request from PublicationsManager@ukaea.uk.

\section*{ORCID iDs}

Teresa Orr: \href{https://orcid.org/0009-0001-9193-0830}{0009-0001-9193-0830}\\
Ayobami Daramola: \href{https://orcid.org/0000-0001-9064-9871}{0000-0001-9064-9871}\\
Rémi Delaporte-Mathurin: \href{https://orcid.org/0000-0003-1064-8882}{0000-0003-1064-8882}\\
James Dark: \href{https://orcid.org/0000-0002-0456-7210}{0000-0002-0456-7210}\\
Graeme Ackland: \href{https://orcid.org/0000-0002-1205-7675}{0000-0002-1205-7675}\\
Prashanth Srinivasan: \href{https://orcid.org/0000-0002-9199-4340}{0000-0002-9199-4340}\\
Andrew P. Horsfield: \href{https://orcid.org/0000-0003-4533-666X}{0000-0003-4533-666X}\\
Mikhail Yu. Lavrentiev: \href{https://orcid.org/0000-0001-9899-1700}{0000-0001-9899-1700}\\
Duc Nguyen-Manh: \href{https://orcid.org/0000-0001-6061-9946}{0000-0001-6061-9946}\\
Mark R. Gilbert: \href{https://orcid.org/0000-0001-8935-1744}{0000-0001-8935-1744}\\
Matthew R. Ryder: \href{https://orcid.org/0000-0002-1363-8148}{0000-0002-1363-8148}

\bibliographystyle{unsrtnat}
\bibliography{references}

@article{abe2008jupiter,
  author  = {Abe, K. and Kohyama, A. and Tanaka, S. and Namba, C. and Terai, T. and Kunugi, T. and Muroga, T. and Hasegawa, A. and Sagara, A. and Berk, S. and Zinkle, S. J. and Sze, D. K. and Petti, D. A. and Abdou, M. A. and Morley, N. B. and Kurtz, R. J. and Snead, L. L. and Ghoniem, N. M.},
  title   = {Development of advanced blanket performance under irradiation and system integration through {JUPITER-II} project},
  journal = {Fusion Engineering and Design},
  volume  = {83},
  number  = {7--9},
  pages   = {842--849},
  year    = {2008},
  doi     = {10.1016/j.fusengdes.2008.07.028}
}

@article{anderl2004deuterium,
  author  = {Anderl, R. A. and Fukada, S. and Smolik, G. R. and Pawelko, R. J. and Schuetz, S. T. and Sharpe, J. P. and Merrill, B. J. and Petti, D. A. and Nishimura, H. and Terai, T. and Tanaka, S.},
  title   = {Deuterium/tritium behavior in {FLiBe} and {FLiBe}-facing materials},
  journal = {Journal of Nuclear Materials},
  volume  = {329--333},
  pages   = {1327--1331},
  year    = {2004},
  doi     = {10.1016/j.jnucmat.2004.04.220}
}

@article{baral2021temperature,
  author  = {Baral, Khagendra and San, Saro and Sakidja, Ridwan and Couet, Adrien and Sridharan, Kumar and Ching, Wai-Yim},
  title   = {Temperature-dependent properties of molten {Li2BeF4} salt using ab initio molecular dynamics},
  journal = {ACS Omega},
  volume  = {6},
  number  = {30},
  pages   = {19822--19835},
  year    = {2021},
  doi     = {10.1021/acsomega.1c02528}
}

@article{calderoni2008permeation,
  author  = {Calderoni, P. and Sharpe, P. and Hara, M. and Oya, Y.},
  title   = {Measurement of tritium permeation in {FLiBe} ({2LiF--BeF2})},
  journal = {Fusion Engineering and Design},
  volume  = {83},
  number  = {7--9},
  pages   = {1331--1334},
  year    = {2008},
  doi     = {10.1016/j.fusengdes.2008.05.016}
}

@article{carotti2021hydrogen,
  author  = {Carotti, Francesco and Liu, Ertai and Macdonald, Digby D. and Scarlat, Raluca O.},
  title   = {An electrochemical study of hydrogen in molten {2LiF--BeF2} ({FLiBe}) with addition of {LiH}},
  journal = {Electrochimica Acta},
  volume  = {367},
  pages   = {137114},
  year    = {2021},
  doi     = {10.1016/j.electacta.2020.137114}
}

@article{delaporteMathurin2024festim,
  author  = {Delaporte-Mathurin, R\'{e}mi and Dark, James and Ferrero, Gabriele and Hodille, Etienne A. and Kulagin, Vladimir and Meschini, Samuele},
  title   = {{FESTIM}: An open-source code for hydrogen transport simulations},
  journal = {International Journal of Hydrogen Energy},
  volume  = {63},
  pages   = {786--802},
  year    = {2024},
  doi     = {10.1016/j.ijhydene.2024.03.184}
}

@article{delaporteMathurin2025baby,
  author  = {Delaporte-Mathurin, R\'{e}mi and Goles, Nikola and Ball, John Leland and Dunn, Collin and Edwards, Emily and Ferry, Sara and Lamere, Edward and Lanzrath, Andrew and Leccacorvi, Rick and Meschini, Samuele and Peterson, Ethan and Segantin, Stefano and Vieira, Rui and Whyte, Dennis and Zhou, Weiyue and Woller, Kevin Benjamin},
  title   = {Advancing tritium self-sufficiency in fusion power plants: Insights from the {BABY} experiment},
  journal = {Nuclear Fusion},
  volume  = {65},
  pages   = {026037},
  year    = {2025},
  doi     = {10.1088/1741-4326/ada2ab}
}

@article{fayfar2024complex,
  author  = {Fayfar, Sean and Chahal, Rajni and Williams, Haley and Gardner, D. Nathanael and Zheng, Guiqiu and Sprouster, David and Neuefeind, J\"{o}rg C. and Olds, Dan and Hwang, Andrea and McFarlane, Joanna and Gallagher, Ryan C. and Asta, Mark and Lam, Stephen and Scarlat, Raluca O. and Khaykovich, Boris},
  title   = {Complex structure of molten {FLiBe} ({2LiF--BeF2}) examined by experimental neutron scattering, {X-ray} scattering, and deep-neural-network based molecular dynamics},
  journal = {PRX Energy},
  volume  = {3},
  pages   = {013001},
  year    = {2024},
  doi     = {10.1103/PRXEnergy.3.013001}
}

@article{ferry2023libra,
  author  = {Ferry, Sara E. and Woller, Kevin B. and Peterson, Ethan E. and Sorensen, Caroline and Whyte, Dennis G.},
  title   = {The {LIBRA} experiment: Investigating robust tritium accountancy in molten {FLiBe} exposed to a {D-T} fusion neutron spectrum},
  journal = {Fusion Science and Technology},
  volume  = {79},
  number  = {1},
  pages   = {13--35},
  year    = {2023},
  doi     = {10.1080/15361055.2022.2078136}
}

@article{field1967solubilities,
  author  = {Field, P. E. and Shaffer, J. H.},
  title   = {The solubilities of hydrogen fluoride and deuterium fluoride in molten fluorides},
  journal = {The Journal of Physical Chemistry},
  volume  = {71},
  number  = {10},
  pages   = {3218--3222},
  year    = {1967},
  doi     = {10.1021/j100869a013}
}

@article{forsberg2017tritiumControl,
  author  = {Forsberg, Charles W. and Lam, Stephen and Carpenter, David M. and Whyte, Dennis G. and Scarlat, Raluca O. and Contescu, Cristian and Wei, Liu and Stempien, John and Blandford, Edward},
  title   = {Tritium control and capture in salt-cooled fission and fusion reactors: Status, challenges, and path forward},
  journal = {Nuclear Technology},
  volume  = {197},
  number  = {2},
  pages   = {119--139},
  year    = {2017},
  doi     = {10.13182/NT16-101}
}

@article{forsberg2020flibeBlankets,
  author  = {Forsberg, Charles and Zheng, Guiqiu and Ballinger, Ronald G. and Lam, Stephen T.},
  title   = {Fusion blankets and fluoride-salt-cooled high-temperature reactors with {FLiBe} salt coolant: Common challenges, tritium control, and opportunities for synergistic development strategies between fission, fusion, and solar salt technologies},
  journal = {Nuclear Technology},
  volume  = {206},
  number  = {11},
  pages   = {1778--1801},
  year    = {2020},
  doi     = {10.1080/00295450.2019.1691400}
}

@article{fukada2002initial,
  author  = {Fukada, S. and Anderl, R. A. and Hatano, Y. and Schuetz, S. T. and Pawelko, R. J. and Petti, D. A. and Smolik, G. R. and Terai, T. and Nishikawa, M. and Tanaka, S. and Sagara, A.},
  title   = {Initial studies of tritium behavior in {FLiBe} and {FLiBe}-facing material},
  journal = {Fusion Engineering and Design},
  volume  = {61--62},
  pages   = {783--788},
  year    = {2002},
  doi     = {10.1016/S0920-3796(02)00204-1}
}

@article{fukada2003flibeD2,
  author  = {Fukada, S. and Anderl, R. A. and Pawelko, R. J. and Smolik, G. R. and Schuetz, S. T. and O'Brien, J. E. and Nishimura, H. and Hatano, Y. and Terai, T. and Petti, D. A. and Sze, D. K. and Tanaka, S.},
  title   = {{FLiBe-D2} permeation experiment and analysis},
  journal = {Fusion Science and Technology},
  volume  = {44},
  number  = {2},
  pages   = {410--414},
  year    = {2003},
  doi     = {10.13182/FST03-A369}
}

@article{fukada2005diffusion,
  author  = {Fukada, Satoshi and Anderl, Robert A. and Sagara, Akio and Nishikawa, Masabumi},
  title   = {Diffusion coefficient of tritium through molten salt {FLiBe} and rate of tritium leak from fusion reactor system},
  journal = {Fusion Science and Technology},
  volume  = {48},
  number  = {1},
  pages   = {666--669},
  year    = {2005},
  doi     = {10.13182/FST05-A1013}
}

@article{fukada2006control,
  author  = {Fukada, S. and Morisaki, A. and Sagara, A. and Terai, T.},
  title   = {Control of tritium in {FFHR-2} self-cooled {FLiBe} blanket},
  journal = {Fusion Engineering and Design},
  volume  = {81},
  number  = {1--7},
  pages   = {477--483},
  year    = {2006},
  doi     = {10.1016/j.fusengdes.2005.09.017}
}

@article{fukada2007reaction,
  author  = {Fukada, S. and Simpson, M. F. and Anderl, R. A. and Sharpe, J. P. and Katayama, K. and Smolik, G. R. and Oya, Y. and Terai, T. and Okuno, K. and Hara, M. and Petti, D. A. and Tanaka, S. and Sze, D.-K. and Sagara, A.},
  title   = {Reaction rate of beryllium with fluorine ion for {FLiBe} redox control},
  journal = {Journal of Nuclear Materials},
  volume  = {367--370},
  pages   = {1190--1196},
  year    = {2007},
  doi     = {10.1016/j.jnucmat.2007.03.218}
}

@article{fuerst2025mstte,
  author  = {Fuerst, Thomas F. and Bowers, Anthony G. and Gietl, Hanns A. and France, Nicole L. and Loftus, L. Shayne and Riet, Adriaan A. and Eklund, Matthew D. and Taylor, Chase N. and Shimada, Masashi},
  title   = {The molten salt tritium transport experiment: A pumped fluoride salt loop for hydrogen isotope experimentation},
  journal = {Annals of Nuclear Energy},
  volume  = {223},
  pages   = {111659},
  year    = {2025},
  doi     = {10.1016/j.anucene.2025.111659}
}

@article{guo2018corrosion,
  author  = {Guo, Shaoqiang and Zhang, Jinsuo and Wu, Wei and Zhou, Wentao},
  title   = {Corrosion in the molten fluoride and chloride salts and materials development for nuclear applications},
  journal = {Progress in Materials Science},
  volume  = {97},
  pages   = {448--487},
  year    = {2018},
  doi     = {10.1016/j.pmatsci.2018.05.003}
}

@article{hara2006interactions,
  author  = {Hara, M. and Hatano, Y. and Simpson, M. F. and Smolik, G. R. and Sharpe, J. P. and Oya, Y. and Okuno, K. and Nishikawa, M. and Terai, T. and Tanaka, S. and Anderl, R. A. and Petti, D. A. and Sze, D. K.},
  title   = {Interactions between molten {FLiBe} and metallic {Be}},
  journal = {Fusion Engineering and Design},
  volume  = {81},
  number  = {1--7},
  pages   = {561--566},
  year    = {2006},
  doi     = {10.1016/j.fusengdes.2005.06.372}
}

@techreport{humrickhouse2020tritiumTransport,
  author      = {Humrickhouse, Paul W. and Fuerst, Thomas F.},
  title       = {Tritium transport phenomena in molten-salt reactors},
  institution = {Idaho National Laboratory},
  number      = {INL/EXT-20-59927-Rev000},
  year        = {2020},
  url         = {https://www.osti.gov/biblio/1777267}
}

@article{lam2021hydrogenValence,
  author  = {Lam, Stephen T. and Li, Qing-Jie and Mailoa, Jonathan and Forsberg, Charles and Ballinger, Ronald and Li, Ju},
  title   = {The impact of hydrogen valence on its bonding and transport in molten fluoride salts},
  journal = {Journal of Materials Chemistry A},
  volume  = {9},
  number  = {3},
  pages   = {1784--1794},
  year    = {2021},
  doi     = {10.1039/D0TA10576G}
}

@article{li2022hfDiffusion,
  author  = {Li, X. and Xu, T. and Liu, M. and Zuo, Y.},
  title   = {Diffusion behaviors of {HF} in molten {LiF--BeF2} and {LiF--NaF--KF} eutectics studied by {FPMD} simulations and electrochemical techniques},
  journal = {Journal of Nuclear Materials},
  volume  = {572},
  pages   = {154031},
  year    = {2022},
  doi     = {10.1016/j.jnucmat.2022.154031}
}

@article{liu2017molecular,
  author  = {Liu, Shuting and Su, Tao and Cheng, Jinhui and An, Xuehui and Zhang, Peng and Liu, Hongtao and Yao, Side and Xie, Leidong and Hou, Huiqi},
  title   = {Investigation on molecular structure of molten {Li2BeF4} ({FLiBe}) salt by infrared absorption spectra and density functional theory ({DFT})},
  journal = {Journal of Molecular Liquids},
  volume  = {242},
  pages   = {1052--1057},
  year    = {2017},
  doi     = {10.1016/j.molliq.2017.07.051}
}

@article{malinauskas1974solubilities,
  author  = {Malinauskas, A. P. and Richardson, D. M.},
  title   = {The solubilities of hydrogen, deuterium, and helium in molten {Li2BeF4}},
  journal = {Industrial \& Engineering Chemistry Fundamentals},
  volume  = {13},
  number  = {3},
  pages   = {242--245},
  year    = {1974},
  doi     = {10.1021/i160051a015}
}

@techreport{meier2014blankets,
  author      = {Meier, W. R.},
  title       = {Assessment of Tritium Breeding Blankets from a Systems Perspective: Status Report},
  institution = {Lawrence Livermore National Laboratory},
  number      = {LLNL-TR-658973},
  year        = {2014},
  doi         = {10.2172/1165762}
}

@article{moriyama1987chemical,
  author  = {Moriyama, H. and Maeda, S. and Ohmura, T. and Moritani, K. and Oishi, J.},
  title   = {Chemical behaviors of tritium formed in a {LiF--BeF2} mixture},
  journal = {Journal of Nuclear Materials},
  volume  = {148},
  number  = {2},
  pages   = {211--216},
  year    = {1987},
  doi     = {10.1016/0022-3115(87)90115-2}
}

@article{nishimura2001chemical,
  author  = {Nishimura, H. and Suzuki, A. and Terai, T. and Yamawaki, M. and Tanaka, S. and Sagara, A. and Motojima, O.},
  title   = {Chemical behavior of {Li2BeF4} molten salt as a liquid tritium breeder},
  journal = {Fusion Engineering and Design},
  volume  = {58--59},
  pages   = {667--672},
  year    = {2001},
  doi     = {10.1016/S0920-3796(01)00581-6}
}

@article{oishi1989tritiumRecovery,
  author  = {Oishi, Jun and Moriyama, Hirotake and Maeda, Seiichiro and Asaoka, Yoshiyuki},
  title   = {Tritium recovery from molten {LiF--BeF2} salt},
  journal = {Fusion Engineering and Design},
  volume  = {8},
  pages   = {317--321},
  year    = {1989},
  doi     = {10.1016/S0920-3796(89)80124-3}
}

@article{petti2006jupiter,
  author  = {Petti, D. A. and Smolik, G. R. and Simpson, M. F. and Sharpe, J. P. and Anderl, R. A. and Fukada, S. and Hatano, Y. and Hara, M. and Oya, Y. and Terai, T. and Sze, D. K. and Tanaka, S.},
  title   = {{JUPITER-II} molten salt {FLiBe} research: An update on tritium, mobilization and redox chemistry experiments},
  journal = {Fusion Engineering and Design},
  volume  = {81},
  number  = {8--14},
  pages   = {1439--1449},
  year    = {2006},
  doi     = {10.1016/j.fusengdes.2005.08.101}
}

@techreport{pint2024materials,
  author      = {Pint, Bruce A. and Kiosidou, Evangelia and Sulejmanovic, Dino and Humrickhouse, Paul},
  title       = {Materials Assessment for {FLiBe} Fusion Blankets},
  institution = {Oak Ridge National Laboratory},
  number      = {ORNL/TM-2024/3508},
  year        = {2024},
  doi         = {10.2172/2455087},
  url         = {https://www.osti.gov/biblio/2455087}
}

@article{romberger1972liquidus,
  author  = {Romberger, K. A. and Braunstein, J. and Thoma, R. E.},
  title   = {New electrochemical measurements of the liquidus in the lithium fluoride-beryllium fluoride system. Congruency of lithium beryllium fluoride ({Li2BeF4})},
  journal = {The Journal of Physical Chemistry},
  volume  = {76},
  number  = {8},
  pages   = {1154--1159},
  year    = {1972},
  doi     = {10.1021/j100652a012}
}

@article{sawan2003transmutation,
  author  = {Sawan, M. E. and Sze, D.-K.},
  title   = {Transmutation and production rates of elements in {FLiBe} and {FLiNaBe} with impact on chemistry control},
  journal = {Fusion Science and Technology},
  volume  = {44},
  number  = {1},
  pages   = {64--68},
  year    = {2003},
  doi     = {10.13182/FST03-A311}
}

@techreport{shaffer1971preparation,
  author      = {Shaffer, J. H.},
  title       = {Preparation and Handling of Salt Mixtures for the Molten Salt Reactor Experiment},
  institution = {Oak Ridge National Laboratory},
  number      = {ORNL-4616},
  year        = {1971},
  doi         = {10.2172/4074869}
}

@article{simpson2006redox,
  author  = {Simpson, M. F. and Smolik, G. R. and Sharpe, J. P. and Anderl, R. A. and Petti, D. A. and Hatano, Y. and Hara, M. and Oya, Y. and Fukada, S. and Tanaka, S. and Terai, T. and Sze, D. K.},
  title   = {Quantitative measurement of beryllium-controlled redox of hydrogen fluoride in molten {FLiBe}},
  journal = {Fusion Engineering and Design},
  volume  = {81},
  number  = {1--7},
  pages   = {541--547},
  year    = {2006},
  doi     = {10.1016/j.fusengdes.2005.08.031}
}

@article{stempien2016trident,
  author  = {Stempien, John D. and Ballinger, Ronald G. and Forsberg, Charles W.},
  title   = {An integrated model of tritium transport and corrosion in fluoride-salt-cooled high-temperature reactors ({FHRs}), Part {I}: Theory and benchmarking},
  journal = {Nuclear Engineering and Design},
  volume  = {310},
  pages   = {258--272},
  year    = {2016},
  doi     = {10.1016/j.nucengdes.2016.10.051}
}

@article{suzuki1998htRelease,
  author  = {Suzuki, A. and Terai, T. and Tanaka, S.},
  title   = {In-situ {HT} release behavior from molten {Li2BeF4} salt},
  journal = {Fusion Engineering and Design},
  volume  = {39--40},
  pages   = {781--785},
  year    = {1998},
  doi     = {10.1016/S0920-3796(98)00381-0}
}

@article{suzuki1998speciesChange,
  author  = {Suzuki, A. and Terai, T. and Tanaka, S.},
  title   = {Change of tritium species in {Li2BeF4} molten salt breeder under neutron irradiation at elevated temperature},
  journal = {Journal of Nuclear Materials},
  volume  = {258--263},
  pages   = {519--524},
  year    = {1998},
  doi     = {10.1016/S0022-3115(98)00423-1}
}

@article{suzuki1998mechanism,
  author  = {Suzuki, A. and Terai, T. and Tanaka, S.},
  title   = {Mechanism on change of tritium species in {Li2BeF4} molten salt breeder under neutron irradiation at elevated temperature},
  journal = {Fusion Technology},
  volume  = {34},
  number  = {3P2},
  pages   = {526--530},
  year    = {1998},
  doi     = {10.13182/FST98-A11963666}
}

@article{suzuki2000permeation,
  author  = {Suzuki, A. and Terai, T. and Tanaka, S.},
  title   = {Tritium release behavior from {Li2BeF4} molten salt by permeation through structural materials},
  journal = {Fusion Engineering and Design},
  volume  = {51--52},
  pages   = {863--868},
  year    = {2000},
  doi     = {10.1016/S0920-3796(00)00233-7}
}

@article{terai1990ctr,
  author  = {Terai, T.},
  title   = {In-situ tritium release behavior from {CTR} liquid breeding materials under 14 {MeV} neutron irradiation},
  journal = {Journal of Nuclear Science and Technology},
  volume  = {27},
  number  = {7},
  pages   = {667--670},
  year    = {1990},
  doi     = {10.3327/jnst.27.667}
}

@article{terai1996intrex,
  author  = {Terai, T. and Suzuki, A. and Tanaka, S.},
  title   = {In-situ tritium release experiment from molten {Li2BeF4} salt under neutron irradiation at elevated temperatures ({INTREX-FLiBe})},
  journal = {Fusion Technology},
  volume  = {30},
  number  = {3P2A},
  pages   = {911--915},
  year    = {1996},
  doi     = {10.13182/FST96-A11963054}
}

@article{terai2001release,
  author  = {Terai, T. and Suzuki, A. and Tanaka, S.},
  title   = {Tritium release from {Li2BeF4} molten salt breeder under neutron irradiation at elevated temperature},
  journal = {Fusion Technology},
  volume  = {39},
  number  = {2P2},
  pages   = {768--772},
  year    = {2001},
  doi     = {10.13182/FST01-A11963331}
}

@article{vergari2023corrosionActivation,
  author  = {Vergari, Lorenzo and Scarlat, Raluca O. and Hayes, Ryan D. and Fratoni, Massimiliano},
  title   = {The corrosion effects of neutron activation of {2LiF--BeF2} ({FLiBe})},
  journal = {Nuclear Materials and Energy},
  volume  = {34},
  pages   = {101289},
  year    = {2023},
  doi     = {10.1016/j.nme.2022.101289}
}

@article{wang2022tritiumFpmd,
  author  = {Wang, H. and Yue, B. and Yan, L. and Jiang, T. and Peng, S.},
  title   = {First-principles molecular dynamics study of the behavior of tritium in molten {LiF--BeF2} eutectic},
  journal = {Journal of Molecular Liquids},
  volume  = {345},
  pages   = {117027},
  year    = {2022},
  doi     = {10.1016/j.molliq.2021.117027}
}

@article{zhang2018redox,
  author  = {Zhang, Jinsuo and Forsberg, Charles W. and Simpson, Michael F. and Guo, Shaoqiang and Lam, Stephen T. and Scarlat, Raluca O. and Carotti, Francesco and Chan, Kevin J. and Singh, Preet M. and Doniger, William and Sridharan, Kumar and Keiser, James R.},
  title   = {Redox potential control in molten salt systems for corrosion mitigation},
  journal = {Corrosion Science},
  volume  = {144},
  pages   = {44--53},
  year    = {2018},
  doi     = {10.1016/j.corsci.2018.08.035}
}

@article{smith2020phase,
  author  = {Smith, A. L. and Capelli, E. and Konings, R. J. M. and Gheribi, A. E.},
  title   = {A new approach for coupled modelling of the structural and thermo-physical properties of molten salts. {Case} of a polymeric liquid {LiF-BeF2}},
  journal = {Journal of Molecular Liquids},
  volume  = {299},
  pages   = {112165},
  year    = {2020},
  doi     = {10.1016/j.molliq.2019.112165}
}

@article{candido2026sensor,
  author  = {Candido, Luigi and Hinz, Alexander and Barron, Paul and Baus, Colin and McGrady, John and Oliver, Mario and Pearson, Richard and Raeves, Ben and Takamine, Jun and Taylor, Jack and Taylor-King, Luke and Ueguchi, Satoshi and Wilkins, Andrew and Wilson, Andrew and Konishi, Satoshi},
  title   = {Tritium transport modelling and engineering design of a hydrogen isotope permeation sensor for {FLiBe} breeder loops},
  journal = {Fusion Engineering and Design},
  volume  = {227},
  pages   = {115728},
  year    = {2026},
  doi     = {10.1016/j.fusengdes.2026.115728}
}

@article{saraswat2026permeation,
  author  = {Saraswat, Abhishek and Zhou, Weiyue and Kim, Nayoung and Cota, Jaron F. and Zheng, Guiqiu and Khechfe, Alexander A. and Barthel, Caroline S. and Short, Michael P. and Delaporte-Mathurin, R\'{e}mi and Woller, Kevin B.},
  title   = {Permeation behaviour of hydrogen isotopes in molten {FLiBe} ({2LiF-BeF2}): Identifying sources of uncertainty and associated measurement challenges},
  journal = {International Journal of Hydrogen Energy},
  volume  = {253},
  pages   = {156016},
  year    = {2026},
  doi     = {10.1016/j.ijhydene.2026.156016}
}

@article{jiang2026interfaces,
  author  = {Jiang, Linbing and Lin, Shuying and Zhou, Xiaobin and Deng, Ke and Yang, Guo and Zeng, Youshi and Liu, Wenguan},
  title   = {Atomic-scale insights into tritium speciation and interfacial behavior at molten salt/graphite and molten salt/alloy interfaces},
  journal = {International Journal of Hydrogen Energy},
  volume  = {252},
  pages   = {156188},
  year    = {2026},
  doi     = {10.1016/j.ijhydene.2026.156188}
}

@article{park2026crossScale,
  author  = {Park, Wooseong and Su, Guanyu and Jeong, Yong Hoon},
  title   = {Integrated cross-scale modeling of dissolved phase transport in molten salt systems},
  journal = {International Journal of Heat and Mass Transfer},
  volume  = {264},
  pages   = {128777},
  year    = {2026},
  doi     = {10.1016/j.ijheatmasstransfer.2026.128777}
}

@article{li2026gasLiquidExtraction,
  author  = {Li, Benjamin James and Su, Guanyu and Dai, Dezhi and Yuan, Haomin and Saini, Nadish},
  title   = {Toward efficient tritium extraction: High-fidelity simulation of gas-liquid contactors in molten salts},
  journal = {International Journal of Heat and Mass Transfer},
  volume  = {267},
  pages   = {128915},
  year    = {2026},
  doi     = {10.1016/j.ijheatmasstransfer.2026.128915}
}

@article{klix2006tritium,
  author  = {Klix, A. and Suzuki, A. and Terai, T.},
  title   = {Study of tritium migration in liquid {Li2BeF4} with ab initio molecular dynamics},
  journal = {Fusion Engineering and Design},
  volume  = {81},
  number  = {1--7},
  pages   = {713--717},
  year    = {2006},
  doi     = {10.1016/j.fusengdes.2005.09.034}
}

@article{edao2009release,
  author  = {Edao, Yuki and Fukada, Satoshi and Noguchi, Hidetaka and Sagara, Akio},
  title   = {Tritium release from neutron-irradiated {FLiBe} purged out by {Ar-H2} or {Ar} at controlled temperature},
  journal = {Fusion Science and Technology},
  volume  = {55},
  number  = {2},
  pages   = {140--151},
  year    = {2009},
  doi     = {10.13182/FST09-A4067}
}

@article{dolan2020graphite,
  author  = {Dolan, Kieran and Zheng, Guiqiu and Carpenter, David and Huang, Steven and Hu, Lin-Wen},
  title   = {Tritium content and chemical form in nuclear graphite from molten fluoride salt irradiations},
  journal = {Fusion Science and Technology},
  volume  = {76},
  number  = {4},
  pages   = {398--403},
  year    = {2020},
  doi     = {10.1080/15361055.2020.1712993}
}

@article{dolan2021tritium,
  author  = {Dolan, Kieran and Zheng, Guiqiu and Sun, Kaichao and Carpenter, David and Hu, Lin-Wen},
  title   = {Tritium generation, release, and retention from in-core fluoride salt irradiations},
  journal = {Progress in Nuclear Energy},
  volume  = {131},
  pages   = {103576},
  year    = {2021},
  doi     = {10.1016/j.pnucene.2020.103576}
}

@article{ferrero2022arc,
  author  = {Ferrero, Gabriele and Meschini, Samuele and Testoni, Raffaella},
  title   = {A preliminary {CFD} and tritium transport analysis for {ARC} blanket},
  journal = {Fusion Science and Technology},
  volume  = {78},
  number  = {8},
  pages   = {617--630},
  year    = {2022},
  doi     = {10.1080/15361055.2022.2096365}
}

@article{dolan2023experimental,
  author  = {Dolan, Kieran and Su, Guanyu and Zheng, Guiqiu and Ames, Michael and Carpenter, David and Hu, Lin-Wen},
  title   = {Experimental measurement and multiphysics simulation of tritium transport in neutron-irradiated {FLiBe} salt},
  journal = {Nuclear Technology},
  volume  = {209},
  number  = {4},
  pages   = {515--531},
  year    = {2023},
  doi     = {10.1080/00295450.2022.2135933}
}

@article{jiang2025solute,
  author  = {Jiang, Linbing and Zeng, Youshi and Deng, Ke and Yang, Guo and Yin, Huiqin and Qiu, Jie and Wu, Xijun and Liu, Wenguan},
  title   = {Solute-driven modulation of tritium behavior in {FLiBe} molten salts: A machine learning molecular dynamics approach},
  journal = {International Journal of Hydrogen Energy},
  volume  = {174},
  pages   = {151411},
  year    = {2025},
  doi     = {10.1016/j.ijhydene.2025.151411}
}

@article{dark2026festimV2,
  author  = {Dark, James and Delaporte-Mathurin, R\'{e}mi and Dokken, J{\o}rgen S. and Yang, Huihua and Khurana, Chirag and Dunnell, Kaelyn and Ferrero, Gabriele and Kulagin, Vladimir and Meschini, Samuele and Dufour, Jonathan and Hodille, Etienne A.},
  title   = {{FESTIM} v2.0: Upgraded framework for multi-species hydrogen transport and enhanced performance},
  journal = {International Journal of Hydrogen Energy},
  volume  = {220},
  pages   = {153987},
  year    = {2026},
  doi     = {10.1016/j.ijhydene.2026.153987}
}

@article{delaporteMathurin2026baby1l,
  author  = {Delaporte-Mathurin, R\'{e}mi and Goles, Nikola and Dunn, Collin and Edwards, Emily and Ferry, Sara and MacDonald, Ross and Peterson, Ethan and Pettinari, Davide and Segantin, Stefano and Zhou, Weiyue and Woller, Kevin B.},
  title   = {{BABY 1L}: First tritium breeding campaign results},
  journal = {Nuclear Fusion},
  volume  = {66},
  pages   = {076046},
  year    = {2026},
  doi     = {10.1088/1741-4326/ae6f31}
}

@misc{yang2026multidimensional,
  author       = {Yang, Huihua and Saraswat, Abhishek and Zhou, Weiyue and Woller, Kevin and Dark, James and Khurana, Chirag and Dunnell, Kaelyn and Peterson, Ethan and Delaporte-Mathurin, R\'{e}mi},
  title        = {Quantifying multidimensional transport effects on permeability inference in {FLiBe} systems using a validation-informed modeling framework},
  year         = {2026},
  howpublished = {arXiv preprint},
  note         = {arXiv:2605.12750},
  doi          = {10.48550/arXiv.2605.12750}
}

@misc{dark2026arcPreprint,
  author       = {Dark, James and Sircar, Arpan and Bae, Jin Whan and Dokken, J{\o}rgen S. and Delaporte-Mathurin, R\'{e}mi},
  title        = {Multiphysics tritium transport modelling of the {ARC} breeding blanket with {FESTIM}},
  year         = {2026},
  howpublished = {arXiv preprint},
  note         = {arXiv:2608.05398},
  doi          = {10.48550/arXiv.2608.05398}
}

@article{bowers2026mstteCFD,
  author  = {Bowers, Jr., Anthony G. and Sharma, Subash L. and Fuerst, Thomas F.},
  title   = {Computational fluid dynamics analysis of the molten salt tritium transport experiment test section},
  journal = {Fusion Science and Technology},
  volume  = {82},
  number  = {1--2},
  pages   = {431--448},
  year    = {2026},
  doi     = {10.1080/15361055.2025.2549608}
}

@article{nakamura2015hydrogen,
  author  = {Nakamura, Akira and Fukada, Satoshi and Nishiumi, Ryosuke},
  title   = {Hydrogen isotopes permeation in a fluoride molten salt for nuclear fusion blanket},
  journal = {Journal of Plasma and Fusion Research SERIES},
  volume  = {11},
  pages   = {25--29},
  year    = {2015},
  url     = {https://www.jspf.or.jp/JPFRS/PDF/Vol11/jpfrs2015_11-025.pdf}
}

@article{nishiumi2016hydrogen,
  author  = {Nishiumi, Ryosuke and Fukada, Satoshi and Nakamura, Akira and Katayama, Kazunari},
  title   = {Hydrogen permeation through {FLiNaBe} fluoride molten salts for blanket candidates},
  journal = {Fusion Engineering and Design},
  volume  = {109--111},
  pages   = {1663--1668},
  year    = {2016},
  doi     = {10.1016/j.fusengdes.2015.10.035}
}

@article{armstrong2020dynamic,
  author  = {Armstrong, Jeff and O'Malley, Alexander J. and Ryder, Matthew R. and Butler, Keith T.},
  title   = {Understanding dynamic properties of materials using neutron spectroscopy and atomistic simulation},
  journal = {Journal of Physics Communications},
  volume  = {4},
  number  = {7},
  pages   = {072001},
  year    = {2020},
  doi     = {10.1088/2399-6528/ab9c2e}
}

@article{barker2025flibe,
  author  = {Barker, Adam J. and Gilbert, Mark R. and Edmondson, Philip D.},
  title   = {Analysis of a segmentation approach to breeder blanket design and the utilisation of {FLiBe} as a novel neutron reflector},
  journal = {Nuclear Fusion},
  volume  = {65},
  number  = {8},
  pages   = {086022},
  year    = {2025},
  doi     = {10.1088/1741-4326/adecb2}
}

@article{gilbert2026librti,
  author  = {Gilbert, Mark R. and Foster, David and Bradnam, Steven C. and Davis, Andrew and Martis, Vladimir and Lavrentiev, Mikhail Yu.},
  title   = {Modeling of a Neutron Source Test Bed for the Fusion Fuel Cycle},
  journal = {Nuclear Science and Engineering},
  year    = {2026},
  doi     = {10.1080/00295639.2026.2652764}
}

@inproceedings{brooks2022blankets,
  author    = {Brooks, Helen and Dixon, Stephen and Davis, Andrew},
  title     = {Towards Multiphysics Simulations of Fusion Breeder Blankets},
  booktitle = {International Conference on Physics of Reactors 2022 (PHYSOR 2022)},
  pages     = {2480--2489},
  year      = {2022},
  publisher = {American Nuclear Society},
  doi       = {10.13182/PHYSOR22-37726}
}

@article{smith2020erratum,
  author  = {Smith, A. L. and Capelli, E. and Konings, R. J. M. and Gheribi, A. E.},
  title   = {Erratum to: ``{A} new approach for coupled modelling of the structural and thermo-physical properties of molten salts. {Case} of a polymeric liquid {LiF-BeF2}''},
  journal = {Journal of Molecular Liquids},
  volume  = {314},
  pages   = {113421},
  year    = {2020},
  doi     = {10.1016/j.molliq.2020.113421}
}

@article{carpenter2017partitioning,
  author  = {Carpenter, David and Ames, Michael and Zheng, Guiqiu and Kohse, Gordon and Hu, Lin-Wen},
  title   = {Tritium Production and Partitioning from the Irradiation of Lithium-Beryllium Fluoride Salt},
  journal = {Fusion Science and Technology},
  volume  = {71},
  number  = {4},
  pages   = {549--554},
  year    = {2017},
  doi     = {10.1080/15361055.2017.1291040}
}

@techreport{ornl4548,
  author      = {Rosenthal, M. W. and Briggs, R. B. and Kasten, P. R.},
  title       = {Molten-Salt Reactor Program Semiannual Progress Report for Period Ending {February} 28, 1970},
  institution = {Oak Ridge National Laboratory},
  number      = {ORNL-4548},
  address     = {Oak Ridge, Tennessee},
  month       = aug,
  year        = {1970}
}

@article{zheng2019tpb,
  author  = {Zheng, Guiqiu and Carpenter, David and Dolan, Kieran and Hu, Lin-Wen},
  title   = {Experimental investigation of alumina coating as tritium permeation barrier for molten salt nuclear reactors},
  journal = {Nuclear Engineering and Design},
  volume  = {353},
  pages   = {110232},
  year    = {2019},
  doi     = {10.1016/j.nucengdes.2019.110232}
}

@misc{delaporteMathurin2026kineticInterface,
  author       = {Delaporte-Mathurin, R\'{e}mi and Dark, James},
  title        = {When chemical potential continuity fails: kinetic interface models for hydrogen isotope transport},
  year         = {2026},
  howpublished = {arXiv preprint},
  note         = {arXiv:2608.26231},
  doi          = {10.48550/arXiv.2608.26231}
}

@article{galashev2023structure,
  author  = {Galashev, A. Y. and Anisimov, A. F. and Vorob'ev, A. S.},
  title   = {Structure and Kinetic Properties of a Molten {FLiBe} Mixture with Tritium},
  journal = {Russian Journal of Physical Chemistry A},
  volume  = {97},
  number  = {12},
  pages   = {2656--2664},
  year    = {2023},
  doi     = {10.1134/S0036024423120099}
}

@article{galashev2025firstPrinciples,
  author  = {Galashev, A. E. and Anisimov, A. F.},
  title   = {A Basic Study of Tritium Behavior in Molten {FLiBe}},
  journal = {Russian Journal of Physical Chemistry A},
  volume  = {99},
  number  = {3},
  pages   = {438--445},
  year    = {2025},
  doi     = {10.1134/S0036024424703473}
}

@article{hunsberger2026fluoroacidity,
  author  = {Hunsberger, O. and Rakib, G. S. and Vergari, L.},
  title   = {Acid-base equilibria in molten fluoride salts and their impact on molten salt reactors},
  journal = {Journal of Fluorine Chemistry},
  volume  = {291},
  pages   = {110569},
  year    = {2026},
  doi     = {10.1016/j.jfluchem.2026.110569}
}

@article{fu2026oxide,
  author  = {Fu, Xiaoqin and Wang, Lei and Li, Yingjie and Li, Xu
             and Li, Zhuyao and Huang, Yuyang and Fu, Haiying
             and Dou, Qiang and Peng, Hao},
  title   = {In Situ Electrochemical Monitoring and Removal of
             Oxide Ions in {FLiBe} Molten Salts},
  journal = {Materials},
  volume  = {19},
  number  = {18},
  pages   = {3868},
  year    = {2026},
  doi     = {10.3390/ma19183868}
}

@article{simon2025tmap8,
  author  = {Simon, Pierre-Cl{\'e}ment A. and Icenhour, Casey T. and Singh, Gyanender and Lindsay, Alexander D. and Bhave, Chaitanya V. and Yang, Lin and Riet, Adriaan A. and Che, Yifeng and Humrickhouse, Paul and Calderoni, Pattrick and Shimada, Masashi},
  title   = {{MOOSE}-based Tritium Migration Analysis Program, Version 8 ({TMAP8}) for Advanced Open-Source Tritium Transport and Fuel Cycle Modeling},
  journal = {Fusion Engineering and Design},
  volume  = {214},
  pages   = {114874},
  year    = {2025},
  doi     = {10.1016/j.fusengdes.2025.114874}
}

@article{pint2026compatibility,
  author  = {Pint, B. A. and {De Lamater-Brotherton}, C. and Sulejmanovic, D. and Lance, M. J.},
  title   = {{FLiBe} compatibility of fusion-relevant materials},
  journal = {Journal of Nuclear Materials},
  volume  = {632},
  pages   = {156925},
  year    = {2026},
  doi     = {10.1016/j.jnucmat.2026.156925}
}

@article{ko2021chargeTransfer,
  author  = {Ko, Tsz Wai and Finkler, Jonas A. and Goedecker, Stefan and Behler, J{\"o}rg},
  title   = {A fourth-generation high-dimensional neural network potential with accurate electrostatics including non-local charge transfer},
  journal = {Nature Communications},
  volume  = {12},
  pages   = {398},
  year    = {2021},
  doi     = {10.1038/s41467-020-20427-2}
}

@article{evans2026sparging,
  author  = {Evans, Anthony and Rakib, G. S. and Vergari, L.},
  title   = {Physics and engineering review of gas sparging in molten salt reactors},
  journal = {Nuclear Engineering and Design},
  volume  = {446},
  pages   = {114624},
  year    = {2026},
  doi     = {10.1016/j.nucengdes.2025.114624}
}

@article{shi2022moltenThermodynamics,
  author  = {Shi, Yu and Lam, Stephen T. and Beck, Thomas L.},
  title   = {Deep neural network based quantum simulations and quasichemical theory for accurate modeling of molten salt thermodynamics},
  journal = {Chemical Science},
  volume  = {13},
  number  = {28},
  pages   = {8265--8273},
  year    = {2022},
  doi     = {10.1039/D2SC02227C}
}

@misc{das2026quantum,
  author       = {Das, Susanta and Santos, Thiago J. Pinheiro Dos
                  and Bhowmik, Subhamoy and Bazayeva, Milana
                  and Li, Zhen and Shajan, Akhil and Kaliakin, Danil
                  and Liang, Fangchun and Bryantsev, Vyacheslav S.
                  and Geist, Al and Gomez, Abigail McClain
                  and Pellegrini, Thaddeus and Walkup, Robert
                  and Seelam, Seetharami R. and Motta, Mario
                  and Merz, Jr., Kenneth M. and Beck, Thomas},
  title        = {Quantum Computations on Fusion Blanket Molten Salts},
  year         = {2026},
  howpublished = {arXiv preprint},
  note         = {arXiv:2606.30402},
  doi          = {10.48550/arXiv.2606.30402}
}

@article{li2021raman,
  author  = {Li, Yangjuan and Liu, Xiyan and Wang, Baozhu and Wang, Chenyang},
  title   = {Raman and theoretical studies on structural evolution of {Li$_2$BeF$_4$} and binary {LiF}--{BeF$_2$} melts},
  journal = {Journal of Molecular Liquids},
  volume  = {325},
  pages   = {115208},
  year    = {2021},
  doi     = {10.1016/j.molliq.2020.115208}
}

@article{fu2023probing,
  author  = {Fu, Xiaobin and Liu, Yiyang and Huang, Hailong and Wu, Huiyan and Sun, Jianchao and Han, Ling and Ge, Min and Qian, Yuan and Liu, Hongtao},
  title   = {Probing the local structure of {FLiBe} melts and solidified salts by in situ high-temperature {NMR}},
  journal = {Physical Chemistry Chemical Physics},
  volume  = {25},
  number  = {29},
  pages   = {19446--19452},
  year    = {2023},
  doi     = {10.1039/D3CP01096A}
}

@article{Daramola2026FLiBe,
  author  = {Daramola, Ayobami and Martis, Vladimir and Ackland, Graeme J. and Nguyen-Manh, Duc},
  title   = {Solubility and Diffusion of Hydrogen Isotopes and Helium in Molten {FLiBe}},
  journal = {arXiv preprint arXiv:2611.20801},
  year    = {2026},
  doi     = {10.48550/arXiv.2611.20801}
}

\end{document}